\documentclass[ijoc,nonblindrev]{informs3} 

\OneAndAHalfSpacedXII 

\usepackage{algorithm}
\usepackage{algpseudocode}
\usepackage{multicol}
\usepackage{multirow}
\usepackage[FIGTOPCAP]{subfigure}

\usepackage[square, numbers]{natbib}
\usepackage{natbib}
 \bibpunct[, ]{(}{)}{,}{a}{}{,}%
 \def\bibfont{\small}%
\TheoremsNumberedThrough     

\EquationsNumberedThrough    

\begin{document}

\RUNTITLE{Computing Algorithm for an Equilibrium of the Generalized Stackelberg Game}

\TITLE{Computing Algorithm for an Equilibrium \\ of the Generalized Stackelberg Game}

\ARTICLEAUTHORS{
\AUTHOR{Jaeyeon Jo, Jihwan Yu, Jinkyoo Park\thanks{Corresponding author}}
\AFF{Department of Industrial and Systems Engineering, KAIST, Daejeon, Republic of Korea,\\ \EMAIL{\{robin512, jihwan14, jinkyoo.park\}@kaist.ac.kr}}
}

\ABSTRACT{
The $1-N$ generalized Stackelberg game (single-leader multi-follower game) is intricately intertwined with the interaction between a leader and followers (hierarchical interaction) and the interaction among followers (simultaneous interaction). However, obtaining the optimal strategy of the leader is generally challenging due to the complex interactions among the leader and followers. Here, we propose a general methodology to find a generalized Stackelberg equilibrium of a $1-N$ generalized Stackelberg game. Specifically, we first provide the conditions where a generalized Stackelberg equilibrium always exists using the variational equilibrium concept. Next, to find an equilibrium in polynomial time, we transformed the $1-N$ generalized Stackelberg game into a $1-1$ Stackelberg game whose Stackelberg equilibrium is identical to that of the original. Finally, we propose an effective computation procedure based on the projected implicit gradient descent algorithm to find a Stackelberg equilibrium of the transformed $1-1$ Stackelberg game. We validate the proposed approaches using the two problems of deriving operating strategies for EV charging stations: (1) the first problem is optimizing the one-time charging price for EV users, in which a platform operator determines the price of electricity and EV users determine the optimal amount of charging for their satisfaction; and (2) the second problem is to determine the spatially varying charging price to optimally balance the demand and supply over every charging station.
}%

\KEYWORDS{Stackelberg game, single-leader multiple-follower game, Stackelberg equilibrium, bilevel optimization, implicit differentiation}

\maketitle

\section{Introduction}
This paper addresses a (generalized) Stackelberg game (single-leader multi-follower problem) $\Gamma=\left<\mathbf{F}, f_{\mathrm{L}}, \left(f_{i}\right)_{i\in\mathbf{F}}, \Omega_{\mathrm{L}}, \left(\Omega_{i}\right)_{i\in\mathbf{F}}\right>$ which can be formulated as follows:
\setlength{\arraycolsep}{0.0em}
\begin{eqnarray}
\mathbf{y}^{*} &=& \argmax_{\mathbf{y} \in \Omega_{\mathrm{L}}}{f_{\mathrm{L}}\left(\mathbf{y}, \left(\mathbf{x}_{i}^{*}\left(\mathbf{y}\right)\right)_{i\in\mathbf{F}} \right)}\label{eqn: leader for general formulation}\\
\mathbf{x}_{i}^{*} \left(\mathbf{y}\right) &=& \argmax_{\mathbf{x}_{i} \in \Omega_i\left(\mathbf{y}, \mathbf{x}^{*}_{-i}\left(\mathbf{y}\right)\right)}{f_i\left(\mathbf{y}, \mathbf{x}_{i}, \mathbf{x}^{*}_{-i}\left(\mathbf{y}\right)\right)}, \forall i\in\mathbf{F}
\label{eqn: follower for general formulation}
\end{eqnarray}
where $f_{\mathrm{L}}$ be the objective function of the leader, $f_{i}$ be the objective function of the follower $i\in\mathbf{F}$, $\mathbf{y}$ be the leader's decision belonging to their strategy set $\Omega_{\mathrm{L}}$, $\mathbf{x}_{i}$ be the follower $i$'s decision belonging to their strategy set $\Omega_{i}\left(\mathbf{y}, \mathbf{x}_{-i}\right)$, $\mathbf{x}_{-i}=\left(\mathbf{x}_{1}, \cdots, \mathbf{x}_{i-1}, \mathbf{x}_{i+1}, \cdots, \mathbf{x}_{N}\right)$ is the follower's joint decision except follower $i$, and $\mathbf{F}=\left[N\right]:=\left\{1, 2, \cdots, N\right\}$ is a set of followers. In detail, the strategy set of the leader, and the follower $i$ is as: 
\setlength{\arraycolsep}{0.0em}
\begin{eqnarray}
\Omega_{\mathrm{L}} &=& \left\{\mathbf{y}\in\mathbb{R}^{n_{\mathrm{L}}} \middle|
\begin{matrix}
h^{j}_{\mathrm{L}}\left(\mathbf{y}\right) \leq 0, \forall j \in \left[ p_{\mathrm{L}} \right]\\
l^{k}_{\mathrm{L}}\left(\mathbf{y}\right) = 0, \forall k \in \left[ q_{\mathrm{L}} \right]
\end{matrix} \right\}\label{eqn:leader's constraint general} \\
\Omega_{i}\left(\mathbf{y}, \mathbf{x}_{-i}\right) &=& \left\{\mathbf{x}_{i}\in\mathbb{R}^{n_{i}} \middle|
\begin{matrix}
h^{j}_{i}\left(\mathbf{y}, \mathbf{x}_{i}, \mathbf{x}_{-i}\right) \leq 0, \forall j\in\left[ p_i \right]\\
l^{k}_{i}\left(\mathbf{y}, \mathbf{x}_{i}, \mathbf{x}_{-i}\right) = 0, \forall k\in\left[ q_i \right]
\end{matrix}  \right\}\label{eqn:follower's constraint general}
\end{eqnarray}
where $\left[n\right] \triangleq \left\{1, \cdots, n\right\}$, $p_{\mathrm{L}}$ is the number of inequality constraints of the leader, $q_{\mathrm{L}}$ is the number of equality constraints of the leader, $p_{i}$ is the number of inequality constraints of the follower $i$, and $q_{i}$ is the number of equality constraints of the follower $i$. Then, the optimal solution $\left(\mathbf{y}^{*}, \left(\mathbf{x}_{i}^{*}\left( \mathbf{y}^{*} \right)\right)_{i\in\mathbf{F}}\right) \in \Omega_{\mathrm{L}} \times \prod_{i\in\mathbf{F}}{\Omega_{i}\left(\mathbf{y}^{*},\mathbf{x}^{*}_{-i}\left(\mathbf{y}^{*}\right)\right)}$ for equations (\ref{eqn: leader for general formulation}) and (\ref{eqn: follower for general formulation}), is said to be a (generalized) Stackelberg equilibrium \citep{Stackelberg1952Theory}, in which neither leader nor followers have no incentives to deviate unilaterally.

\begin{figure}[t]
\begin{center}
\includegraphics[width=0.95\linewidth]{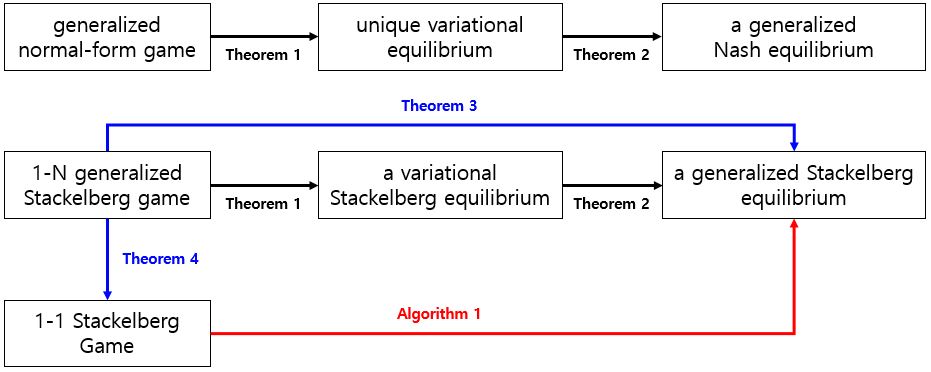}
\caption{Overall procedure to find a generalized Stackelberg equilibrium of the generalized Stackelberg game}
\label{fig:overall}
\end{center}
\end{figure}

The generalized Stackelberg game $\Gamma$ is intricately intertwined with the interaction between a leader and followers (hierarchical interaction) and the interaction among followers (simultaneous interaction). Due to the complex interactions among the leader and followers, computing a generalized Stackelberg equilibrium of the $1-N$ generalized Stackelberg game is challenging. In this study, we propose a general methodology to find a generalized Stackelberg equilibrium of the $1-N$ generalized Stackelberg game (single-leader multi-follower game with followers' joint constraints). First, we provide the conditions where a generalized Stackelberg equilibrium always exists using the variational equilibrium concept. Next, to find an equilibrium in polynomial time, we transform the $1-N$ generalized Stackelberg game into the $1-1$ Stackelberg game whose Stackelberg equilibrium is identical to that of the original. Finally, we propose an effective computation procedure based on the projected implicit gradient descent algorithm to find a Stackelberg equilibrium of the transformed $1-1$ Stackelberg game. Figure \ref{fig:overall} illustrates this general procedure through a schematic diagram composed of theories and an algorithm.

To validate the effectiveness of the proposed modeling framework (i.e., generalized Stackelberg game) and its solution-finding algorithm, we consider two problems of deriving operating strategies for EV charging stations (sharing platform) under the assumption that every EV user will make their decision to minimize their cost. The first is optimizing the one-time charging price for EV users \citep{tushar2012economics}, in which a platform operator determines the price of electricity and EV users determine the optimal amount of charging for their satisfaction. We show the convergence by comparing the results obtained by applying the proposed algorithm to the generalized Stackelberg equilibrium computed analytically. The second problem is to determine the spatially varying charging price to optimally balance the demand and supply over every charging station \citep{zhou2015optimal}. The second problem has a more complex relationship between the leader and the followers and doesn't have any known algorithm to obtain the generalized Stackelberg equilibrium. We compare the performance of the proposed algorithm with the proximal algorithm designed to find a stationary point without considering the hierarchical structure.

The novelty and importance of this study are summarized as follows:
\begin{itemize}
    \item We propose a general methodology to find a generalized Stackelberg equilibrium of the $1-N$ generalized Stackelberg game. First, we provide the conditions where a generalized Stackelberg equilibrium always exists. Next, to find an equilibrium in polynomial time, we develop a method to convert the $1-N$ generalized Stackelberg game into the $1-1$ Stackelberg game whose Stackelberg equilibrium corresponds to a generalized Stackelberg equilibrium of the original game. Finally, we propose a projected implicit gradient descent (PIGD) algorithm to find a Stackelberg equilibrium of the transformed $1-1$ Stackelberg game in polynomial time.
    
    \item We validate the proposed algorithm through EV sharing platforms. We show that our algorithm can always compute a generalized Stackelberg equilibrium of the $1-N$ generalized Stackelberg game. Moreover, we experimentally verified the performance of a generalized Stackelberg equilibrium by comparing its equilibrium value to that of other solution concepts.
\end{itemize}

The organization of the paper is as follows. Section 3 discusses subgame of the generalized Stackelberg game. Section 4 provides the conditions where a generalized Stackelberg equilibrium always exists. Section 5 discusses the computational approach to find a generalized Stackelberg equilibrium. Section 6 introduces the problem description for EV sharing platforms, and Section 7 evaluates the performances of the proposed algorithm using simulation studies.

\section{Backgrounds and Related Works}
The $1-N$ generalized Stackelberg game is intricately intertwined with the hierarchical interaction between a leader and followers and the simultaneous interaction among followers, and these relationships are represented in the forms of bilevel optimization and generalized normal-form game, respectively. While this paper deals with both interactions simultaneously, previous studies have mostly addressed hierarchical interactions in bilevel optimization research and simultaneous interactions in generalized normal-form game research independently.

\subsection{Hierarchical interaction between a leader and a follower}
First, we discuss the hierarchical interaction between a leader and a follower. The problem of hierarchical structure is generally formulated in a kind of a bilevel optimization problem (BOP) \citep{dempe2020bilevel, kleinert2021survey}. A BOP (single-leader single-follower problem) $\Gamma=\left<f_{\mathrm{L}}, f_{\mathbf{F}}, \Omega_{\mathrm{L}}, \Omega_{\mathrm{F}}\right>$ which can be formulated as follows:
\setlength{\arraycolsep}{0.0em}
\begin{eqnarray}
\mathbf{y}^{*} &=& \argmax_{\mathbf{y} \in \Omega_{\mathrm{L}}}{f_{\mathrm{L}}\left(\mathbf{y}, \mathbf{x}^{*}\left(\mathbf{y}\right)\right)}\nonumber\\
\mathbf{x}^{*}\left(\mathbf{y}\right) &=& \argmax_{\mathbf{x} \in \Omega_{\mathrm{F}}}{f_{\mathrm{F}}\left(\mathbf{y}, \mathbf{x}\right)}\label{eqn:bilevel programming general}
\end{eqnarray}
where $f_{\mathrm{L}}$ be the objective function of the leader, $f_{\mathrm{F}}$ be the objective function of a follower, $\mathbf{y}$ be the leader's decision belonging to their strategy set $\Omega_{\mathrm{L}}$, and $\mathbf{x}$ be the follower's decision belonging to their strategy set $\Omega_{\mathrm{F}}$. Then, the optimal solution of $\Gamma$ is defined as $\left(\mathbf{y}^{*}, \mathbf{x}^{*}\left(\mathbf{y}^{*}\right)\right)\in\Omega_{\mathrm{L}}\times\Omega_{\mathrm{F}}$ for equation (\ref{eqn:bilevel programming general}). We classify BOP into two classes: (1) problems where a leader and a follower each have a single objective function; and (2) problems where a leader and a follower have a multi-objective function.

Some studies investigate a method to find the optimal of a BOP consisting of a leader and a follower with a single objective function \citep{mehlitz2021sufficient, contardo2022progressive, gould2021deep}. \cite{mehlitz2021sufficient} suggest first- and second-order sufficient optimality conditions for a solution of BOP. \cite{contardo2022progressive} propose a progressive approximation algorithm and apply it to find the exact solution of a class of interdiction BOP. \cite{gould2021deep} introduce the method to compute the gradient of the follower's decision with respect to the leader's decision and use it to run the implicit gradient descent algorithm. 

Other studies propose an algorithm to compute the optimal solution of the bilevel multi-objective optimization problem (BMOP) consisting of a leader and a follower with more than one objective function 
\citep{dedzo2012solution, lu2014smoothing, chuong2020optimality}. \cite{dedzo2012solution} suggest a new sufficient optimality condition for a solution of a differentiable BMOP. \cite{lu2014smoothing} transform a BMOP into an equivalent nonsmooth multiobjective one-level optimization problem using Karush-Kuhn-Tucker (KKT) conditions. \cite{chuong2020optimality} relaxes a BMOP using the Fritz John (FJ) and KKT conditions to find its optimal solution.

\subsection{Simultaneous interaction among followers}
The modeling of simultaneous interaction among followers is classified depending on whether there are joint constraints among followers, meaning whether a follower's strategy depends on the decisions of other followers. Simultaneous interaction without joint constraints is modeled as a normal-form game and has a Nash equilibrium as a solution \citep{Nash1950Equilibrium}. A normal-form game $\mathrm{G} = \left<\mathbf{F}, \left(f_{i}\right)_{i\in\mathbf{F}}, \left(\Omega_{i}\right)_{i\in\mathbf{F}}\right>$ which can be formulated as follows:
\setlength{\arraycolsep}{0.0em}
\begin{eqnarray}
\mathbf{x}_{i}^{*} &=& \argmax_{\mathbf{x}_{i} \in \Omega_i}{f_i\left(\mathbf{x}_{i}, \mathbf{x}^{*}_{-i}\right)}, \forall i\in\mathbf{F}\label{eqn: normal-form game general form}
\end{eqnarray}
where $f_{i}$ be the objective function of the follower $i\in\mathbf{F}$, $\mathbf{x}_{i}$ be the follower $i$'s decision belonging to their strategy set $\Omega_{i}$, $\mathbf{x}_{-i}=\left(\mathbf{x}_{1}, \cdots, \mathbf{x}_{i-1}, \mathbf{x}_{i+1}, \cdots, \mathbf{x}_{N}\right)$ is the follower's joint decision except follower $i$, and $\mathbf{F}=\left\{1, 2, \cdots, N\right\}$ is a set of followers. Then, the optimal solution $\left(\mathbf{x}_{i}^{*}\right)_{i\in\mathbf{F}}\in\prod_{i\in\mathbf{F}}\Omega_{i}$ for equation (\ref{eqn: normal-form game general form}) is said to be a Nash equilibrium \citep{Nash1950Equilibrium}.

Some studies suggest an algorithm to compute a Nash equilibrium of the normal-form game \citep{tsaknakis2021finding, ahmadi2021semidefinite, bichler2023learning}. \cite{tsaknakis2021finding} propose an algorithm, based on the regularized Nikaido-Isoda (NI) function, for finding the first-order Nash equilibrium of a two-player zero-sum game. \cite{ahmadi2021semidefinite} introduce an semidefinite programming for finding $\epsilon$-approximate Nash equilibrium in bimatrix games. \cite{bichler2023learning} present neural pseudogradient ascent (NPGA) algorithm to compute Bayesian Nash equilibrium in auction games.

In contrast, simultaneous interaction with joint constraints is modeled as a generalized normal-form game and has a generalized Nash equilibrium as a solution \citep{facchinei2007generalized, qu2013methods, ba2022exact}. A generalized normal-form game $\mathrm{G} = \left<\mathbf{F}, \left(f_{i}\right)_{i\in\mathbf{F}}, \left(\Omega_{i}\right)_{i\in\mathbf{F}}\right>$ which can be formulated as follows:
\setlength{\arraycolsep}{0.0em}
\begin{eqnarray}
\mathbf{x}_{i}^{*} &=& \argmax_{\mathbf{x}_{i} \in \Omega_i\left(\mathbf{x}^{*}_{-i}\right)}{f_i\left(\mathbf{x}_{i}, \mathbf{x}^{*}_{-i}\right)}, \forall i\in\mathbf{F}\label{eqn: generalized normal-form game general form}
\end{eqnarray}
where $f_{i}$ be the objective function of the follower $i\in\mathbf{F}$, $\mathbf{x}_{i}$ be the follower $i$'s decision belonging to their strategy set $\Omega_{i}\left(\mathbf{x}_{-i}\right)$, $\mathbf{x}_{-i}=\left(\mathbf{x}_{1}, \cdots, \mathbf{x}_{i-1}, \mathbf{x}_{i+1}, \cdots, \mathbf{x}_{N}\right)$ is the follower's joint decision except follower $i$, and $\mathbf{F}=\left\{1, 2, \cdots, N\right\}$ is a set of followers. Then, the optimal solution $\left(\mathbf{x}_{i}^{*}\right)_{i\in\mathbf{F}}\in\prod_{i\in\mathbf{F}}\Omega_{i}\left(\mathbf{x}_{-i}\right)$ for equation (\ref{eqn: generalized normal-form game general form}) is said to be a generalized Nash equilibrium \citep{facchinei2007generalized}.

Some studies propose an algorithm that computes a generalized Nash equilibrium of the generalized normal-form game where a player's constraints depend on the other player's decisions \citep{von2009optimization, von2009relaxation, zhang2010some}. \cite{von2009optimization} and \cite{von2009relaxation} suggest an algorithm to compute a generalized Nash equilibrium based on the regularized NI function. \cite{zhang2010some} propose two projection-like algorithms for solving a generalized normal-form game.

\subsection{$1-N$ Generalized Stackelberg Game}
There have been studies that use the generalized Stackelberg game concept to model the strategic interactions between a leader and followers and suggest the computational method to compute a generalized Stackelberg equilibrium. Depending on the generality of the problem formulation, the versatility of the problem-solving algorithm is different. We categorize the related studies into: (1) problems that can be solved with the non-deterministic polynomial-time(NP) algorithm, and (2) problems that can be solved with the polynomial-time algorithm.

Some studies propose an NP algorithm for finding a Stackelberg equilibrium of the small-size problem \citep{demiguel2009stochastic, li2018interactive}. \cite{demiguel2009stochastic} model the competition in the telecommunication industry into the Stackelberg game. They compute a stochastic multiple-leader Stackelberg-Nash-Cournot equilibrium by solving mixed-integer non-linear programming (MINLP). \cite{li2018interactive} propose a generalized Stackelberg game modeling the interaction between EVs and power grid operations. This study computed the Stackelberg equilibrium using an MINLP. Although these two studies compute Stackelberg equilibrium using an MINLP solver, they are not adaptable to the complex Stackelberg game since MINLP is an NP algorithm.

Other studies propose a polynomial-time algorithm to find a Stackelberg equilibrium of a problem with an analytically solvable subgame \cite{yoon2015stackelberg, maharjan2013dependable} or a special structure of the objective function \citep{wang2018stackelberg}. \cite{yoon2015stackelberg} propose a Stackelberg game modeling the interaction between retailers and customers who use electric vehicles (EVs). \cite{wang2018stackelberg} modeled the interactions between multiple companies and energy users as an $M-N$ Stackelberg game. By utilizing the property that the subgame can be analytically computed, these studies transform the original Stackelberg game into a normal-form game and compute its equilibrium. However, this method can only be applied to a simple Stackelberg game where the solution of the subgame can be obtained analytically. Wang et al. (2018) modeled an RF-powered cognitive radio network system as the generalized Stackelberg potential game and proposed the directional ascent method to compute the generalized Stackelberg equilibrium. Because the utility function of the potential game is independent of the other players' decisions, the generalized Stackelberg equilibrium is computed in polynomial time.

The studies discussed above compute a generalized Stackelberg equilibrium using strong restrictions of the game structure. The NP algorithm is applied only to small-size problems due to its time complexity. The polynomial-time algorithm is applied only to problems with restrictions on the subgame, or the utility functions. However, the current study computes the equilibrium for a Stackelberg game in polynomial time without sacrificing the generalized structure of the game, such as the structures of the subgame and the utility function of followers.

\section{Generalized Nash Equilibrium for the Subgame of a Generalized Stackelberg Game}
This section introduces the equilibrium analysis for a generalized normal-form game $\mathrm{G}\left(\mathbf{y}\right) = \left<\mathbf{F}, \left( f_{i} \right)_{i \in \mathbf{F}}, \left(\Omega_{i}\right)_{i\in\mathbf{F}}\right>$ where $\mathbf{F}=\left[N\right]$ is a set of followers, $f_{i}$ be the objective function of the follower $i\in\mathbf{F}$, $\Omega_{i}$ is the strategy set of follower $i\in\mathbf{F}$. It models the non-cooperative behavior of followers with the joint constraints when the leader's decision is $\mathbf{y}$. This followers' subgame $\mathrm{G}\left(\mathbf{y}\right)$ of the generalized Stackelberg game $\Gamma$ is described in equations (\ref{eqn: follower for general formulation}) and (\ref{eqn:follower's constraint general}). 
\setlength{\arraycolsep}{0.0em}
\begin{eqnarray}
\mathbf{x}_{i}^{*} \left(\mathbf{y}\right) &=& \argmax_{\mathbf{x}_{i} \in \Omega_i\left(\mathbf{y}, \mathbf{x}^{*}_{-i}\left(\mathbf{y}\right)\right)}{f_i\left(\mathbf{y}, \mathbf{x}_{i}, \mathbf{x}^{*}_{-i}\left(\mathbf{y}\right)\right)}, \forall i\in\mathbf{F}
\nonumber\\ \mathrm{where}\quad
\Omega_{i}\left(\mathbf{y}, \mathbf{x}_{-i}\right) &=& \left\{\mathbf{x}_{i}\in\mathbb{R}^{n_{i}} \middle|
\begin{matrix}
h^{j}_{i}\left(\mathbf{y}, \mathbf{x}_{i}, \mathbf{x}_{-i}\right) \leq 0, \forall j\in\left[ p_i \right]\\
l^{k}_{i}\left(\mathbf{y}, \mathbf{x}_{i}, \mathbf{x}_{-i}\right) = 0, \forall k\in\left[ q_i \right]
\end{matrix}  \right\} \label{eqn:subgame}
\end{eqnarray}
where $\mathbf{x}_{i}$ is the decision of follower $i$, and $\mathbf{x}_{-i}=\left(\mathbf{x}_{1}, \cdots, \mathbf{x}_{i-1}, \mathbf{x}_{i+1}, \cdots, \mathbf{x}_{N}\right)$ is the follower's joint decision except follower $i$.

Here, we provide the conditions where a generalized Nash equilibrium, that is the optimal solution of a generalized normal-form game $\mathrm{G}\left(\mathbf{y}\right)$, always exists by using the theorems regarding the existence and uniqueness of the variational equilibrium of the generalized normal-form game. This can be used to provide the equilibrium conditions between a leader and followers. Figure \ref{fig:overall} illustrates the procedure to conclude the existence of a generalized Nash equilibrium and its extension to a generalized Stackelberg game (this will be discussed in Section 4).

We first define a generalized Nash equilibrium \citep{facchinei2007generalized, qu2013methods}. A generalized Nash equilibrium $\left(\mathbf{x}_{i}^{*}\left(\mathbf{y}\right)\right)_{i\in\mathbf{F}}$ is the joint decision where no user has an incentive to deviate from their decision unless other users change their decisions.

\begin{definition}
\label{def1}
The joint decision $\mathbf{x}^{*}\left(\mathbf{y}\right)\in\prod_{i\in\mathbf{F}}{\Omega_i\left(\mathbf{y}, \mathbf{x}^{*}_{-i}\left(\mathbf{y}\right)\right)}$ is a generalized Nash equilibrium of the subgame $\mathrm{G}\left(\mathbf{y}\right) = \left<\mathbf{F}, \left( f_{i} \right)_{i \in \mathbf{F}}, \left(\Omega_{i}\right)_{i\in\mathbf{F}}\right>$ if 
\setlength{\arraycolsep}{0.0em}
\begin{eqnarray}
f_i\left(\mathbf{y}, \mathbf{x}^{*}\left(\mathbf{y}\right)\right)\geq f_i\left(\mathbf{y}, \mathbf{x}_{i}, \mathbf{x}^{*}_{-i}\left(\mathbf{y}\right)\right), \forall \mathbf{x}_{i}\in\Omega_{i}\left(\mathbf{y}, \mathbf{x}_{-i}^{*}\left(\mathbf{y}\right)\right), \forall i\in\mathbf{F}
\label{eqn:def1}
\end{eqnarray}
where $\mathbf{x}^{*}\left(\mathbf{y}\right) = \left(\mathbf{x}^{*}_{i}\left(\mathbf{y}\right)\right)_{i\in\mathbf{F}}$, and $\mathbf{x}^{*}_{-i}\left(\mathbf{y}\right) = \left(\mathbf{x}^{*}_{1}\left(\mathbf{y}\right), \cdots, \mathbf{x}^{*}_{i-1}\left(\mathbf{y}\right), \mathbf{x}^{*}_{i+1}\left(\mathbf{y}\right), \cdots, \mathbf{x}^{*}_{N}\left(\mathbf{y}\right)\right)$.
\end{definition}

Next, we define a variational equilibrium \citep{facchinei2007generalized, qu2013methods} to prove the existence of a generalized Nash equilibrium. A variational equilibrium is defined as a solution of variational inequality. 
\begin{definition}
\label{def2}
The joint decision $\mathbf{x}^{*}\left(\mathbf{y}\right)\in\prod_{i\in\mathbf{F}}{\Omega_i\left(\mathbf{y}, \mathbf{x}^{*}_{-i}\left(\mathbf{y}\right)\right)}$ is a variational equilibrium of the subgame $\mathrm{G}\left(\mathbf{y}\right) = \left<\mathbf{F}, \left( f_{i} \right)_{i \in \mathbf{F}}, \left(\Omega_{i}\right)_{i\in\mathbf{F}}\right>$ if 
\setlength{\arraycolsep}{0.0em}
\begin{eqnarray}
\mathbf{D}\left(\mathbf{y}, \mathbf{x}^{*}\left(\mathbf{y}\right)\right)^{\mathrm{T}}\left(\mathbf{x}^{*}\left(\mathbf{y}\right)-\mathbf{x}\right)\geq 0, \forall \mathbf{x}\in\prod_{i\in\mathbf{F}}{\Omega_{i}\left(\mathbf{y}, \mathbf{x}_{-i}\right)}
\label{eqn:def2}
\end{eqnarray}
where $\mathbf{D}\left(\mathbf{y}, \mathbf{x}^{*}\left(\mathbf{y}\right)\right)=\left(\nabla_{\mathbf{x}_{i}}{f_i\left(\mathbf{y}, \mathbf{x}^{*}\left(\mathbf{y}\right)\right)}\right)_{i\in\mathbf{F}}$ is the gradient of followers' objective,
$\mathbf{x}^{*}\left(\mathbf{y}\right) = \left(\mathbf{x}^{*}_{i}\left(\mathbf{y}\right)\right)_{i\in\mathbf{F}}$, $\mathbf{x}^{*}_{-i}\left(\mathbf{y}\right) = \left(\mathbf{x}^{*}_{1}\left(\mathbf{y}\right), \cdots, \mathbf{x}^{*}_{i-1}\left(\mathbf{y}\right), \mathbf{x}^{*}_{i+1}\left(\mathbf{y}\right), \cdots, \mathbf{x}^{*}_{N}\left(\mathbf{y}\right)\right)$,
$\mathbf{x}=\left(\mathbf{x}_{i}\right)_{i\in\mathbf{F}}$,
and $\mathbf{x}_{-i} = \left(\mathbf{x}_{1}, \cdots, \mathbf{x}_{i-1}, \mathbf{x}_{i+1}, \cdots, \mathbf{x}_{N}\right)$.
\end{definition}

We cannot compute a generalized Nash equilibrium in general situations because each user's strategy set depends on the decisions of other users. In contrast, the variational equilibrium is also a generalized Nash equilibrium under certain conditions. Therefore, we compute the variational equilibrium to find a generalized Nash equilibrium of the generalized normal-form game that satisfies the conditions described in Theorem \ref{thm2}.

Theorem \ref{thm1} provides the condition for the existence and the uniqueness of the variational equilibrium in the generalized normal-form game. Then, we provide the condition of the existence of a generalized Nash equilibrium using the relationship with the variational equilibrium in Theorem \ref{thm2}.

\begin{theorem}
\label{thm1}
Let $\mathbf{D}\left(\cdot\right)=\left(\nabla_{\mathbf{x}_{i}}{f_i\left(\cdot\right)}\right)_{i\in\mathbf{F}}$ is the gradient of followers' objective.
If the following three conditions are satisfied, then the subgame $\mathrm{G}\left(\mathbf{y}\right) = \left<\mathbf{F}, \left( f_{i} \right)_{i \in \mathbf{F}}, \left(\Omega_{i}\right)_{i\in\mathbf{F}}\right>$ has the unique variational equilibrium.
\begin{itemize}
    \item C
    1. $\Omega_i$ is closed and convex for all $i\in\mathbf{F}$
    \item C
    2. $f_i$ is continuous on $\Omega_i$ for all $i\in\mathbf{F}$
    \item C
    3. $-\mathbf{D}\left(\cdot\right)$ is strongly monotone on $\prod_{i\in\mathbf{F}}{\Omega_i}$
\end{itemize}
\end{theorem}

\proof{Proof of Theorem \ref{thm1}} 
It is proven by Theorem 2.3.3 of \citep{facchinei2003finite}.\Halmos
\endproof

Next, we provide the conditions where the variational equilibrium becomes a generalized Nash equilibrium.

\begin{theorem}
\label{thm2}
If the following three conditions are satisfied, then the variational equilibrium of $\mathrm{G}\left(\mathbf{y}\right) = \left<\mathbf{F}, \left( f_{i} \right)_{i \in \mathbf{F}}, \left(\Omega_{i}\right)_{i\in\mathbf{F}}\right>$ is also a generalized Nash equilibrium of $\mathrm{G}\left(\mathbf{y}\right)$.
\begin{itemize}
    \item C
    1. $\Omega_i$ is closed and convex for {all} $i\in\mathbf{F}$
    \item C
    2. $f_i$ is concave $C^{1}$-function on $\Omega_i$ for {all} $i\in\mathbf{F}$
    \item {C}
    3. $\prod_{i\in\mathbf{F}}{\Omega_i}$ is jointly convex
\end{itemize}
\end{theorem}

\proof{Proof of Theorem \ref{thm2}} 
It is proven by Theorem 5 of \citep{facchinei2007generalized}. \Halmos
\endproof

Theorems \ref{thm1} and \ref{thm2} provide the conditions where a generalized Nash equilibrium exists. That is, if a given subgame $\mathrm{G}\left(\mathbf{y}\right)$ satisfies the conditions in Theorem \ref{thm2}, we can find a generalized Nash equilibrium by computing the unique variational equilibrium of the game $\mathrm{G}\left(\mathbf{y}\right)$. These theorems will be utilized in Section 4 to provide the conditions where the equilibrium of a generalized Stackelberg game exists.

\section{Generalized Stackelberg Equilibrium for the Generalized Stackelberg Game}
In this section, we introduce the generalized Stackelberg equilibrium (optimal solution) of a $1-N$ generalized Stackelberg game (single-leader multi-follower game) $\Gamma=\left<\mathbf{F}, f_{\mathrm{L}}, \left(f_{i}\right)_{i\in\mathbf{F}}, \Omega_{\mathrm{L}}, \left(\Omega_{i}\right)_{i\in\mathbf{F}}\right>$, which is formulated in equations (\ref{eqn: leader for general formulation}) - (\ref{eqn:follower's constraint general}), is as follows:
\setlength{\arraycolsep}{0.0em}
\begin{eqnarray}
\mathbf{y}^{*} &=& \argmax_{\mathbf{y} \in \Omega_{\mathrm{L}}}{f_{\mathrm{L}}\left(\mathbf{y}, \left(\mathbf{x}_{i}^{*}\left(\mathbf{y}\right)\right)_{i\in\mathbf{F}} \right)}\nonumber\\
\mathbf{x}_{i}^{*} \left(\mathbf{y}\right) &=& \argmax_{\mathbf{x}_{i} \in \Omega_i\left(\mathbf{y}, \mathbf{x}^{*}_{-i}\left(\mathbf{y}\right)\right)}{f_i\left(\mathbf{y}, \mathbf{x}_{i}, \mathbf{x}^{*}_{-i}\left(\mathbf{y}\right)\right)}, \forall i\in\mathbf{F}\nonumber\\
\mathrm{where} \quad \Omega_{\mathrm{L}} &=& \left\{\mathbf{y}\in\mathbb{R}^{n_{\mathrm{L}}} \middle|
\begin{matrix}
h^{j}_{\mathrm{L}}\left(\mathbf{y}\right) \leq 0, \forall j \in \left[ p_{\mathrm{L}} \right]\\
l^{k}_{\mathrm{L}}\left(\mathbf{y}\right) = 0, \forall k \in \left[ q_{\mathrm{L}} \right]
\end{matrix} \right\}\nonumber\\
\Omega_{i}\left(\mathbf{y}, \mathbf{x}_{-i}\right) &=& \left\{\mathbf{x}_{i}\in\mathbb{R}^{n_{i}} \middle|
\begin{matrix}
h^{j}_{i}\left(\mathbf{y}, \mathbf{x}_{i}, \mathbf{x}_{-i}\right) \leq 0, \forall j\in\left[ p_i \right]\\
l^{k}_{i}\left(\mathbf{y}, \mathbf{x}_{i}, \mathbf{x}_{-i}\right) = 0, \forall k\in\left[ q_i \right]
\end{matrix}  \right\}\label{eqn:Stackelberg game}
\end{eqnarray}
where $f_{\mathrm{L}}$ be the objective function of the leader, $f_{i}$ be the objective function of the follower $i\in\mathbf{F}$, $\mathbf{y}$ be the leader's decision belonging to the strategy set $\Omega_{\mathrm{L}}$, $\mathbf{x}_{i}$ be the follower $i$'s decision belonging to their strategy set $\Omega_{i}\left(\mathbf{y}, \mathbf{x}_{-i}\right)$, and $\mathbf{F}=\left[N\right]$ is a set of followers. Here, we provide the existence conditions of a generalized Stackelberg equilibrium of the $1-N$ generalized Stackelberg game $\Gamma$.

We first define a generalized Stackelberg equilibrium of the $1-N$ generalized Stackelberg game $\Gamma$. A generalized Stackelberg equilibrium $\left(\mathbf{y}^{*}, \left(\mathbf{x}_{i}^{*}\left(\mathbf{y}^{*}\right)\right)_{i\in\mathbf{F}}\right)$ is the joint decision where a leader makes the optimal decision $\mathbf{y}^{*}$ for maximizing its utility function $f_{\mathrm{L}}$, while followers are on the generalized Nash equilibrium of the subgame $\mathrm{G}\left(\mathbf{y}^{*}\right)$ in equation (\ref{eqn:subgame}).

\begin{definition}
\label{def4}
The joint decision $\left(\mathbf{y}^{*}, \mathbf{x}^{*}\left(\mathbf{y}^{*}\right)\right)\in\Omega_{\mathrm{L}}\times\prod_{i\in\mathbf{F}}{\Omega_i\left(\mathbf{y}^{*}, \mathbf{x}^{*}_{-i}\left(\mathbf{y}^{*}\right)\right)}$ is a generalized Stackelberg equilibrium of $\Gamma=\left<\mathbf{F}, f_{\mathrm{L}}, \left(f_{i}\right)_{i\in\mathbf{F}}, \Omega_{\mathrm{L}}, \left(\Omega_{i}\right)_{i\in\mathbf{F}}\right>$ if 
\setlength{\arraycolsep}{0.0em}
\begin{eqnarray}
\sup_{\mathbf{x}^{*}\left(\mathbf{y}^{*}\right)\in \mathrm{GNE}\left(\mathbf{y}^{*}\right)}f_{\mathrm{L}}\left(\mathbf{y}^{*}, \mathbf{x}^{*}\left(\mathbf{y}^{*}\right)\right)&\geq&\inf_{\mathbf{x}^{*}\left(\mathbf{y}\right)\in \mathrm{GNE}\left(\mathbf{y}\right)}f_{\mathrm{L}}\left(\mathbf{y}, \mathbf{x}^{*}\left(\mathbf{y}\right)\right), \forall \mathbf{y} \in \Omega_{\mathrm{L}}
\label{eqn:def4}
\end{eqnarray}
where $\mathrm{GNE}(\mathbf{y})=\left\{ \mathbf{x}^{*} \in \prod_{i\in\mathbf{F}}\Omega_{i}\left(\mathbf{y},\mathbf{x}^{*}_{-i}\right) \middle| f_i\left(\mathbf{y}, \mathbf{x}^{*}\right) \geq f_i\left(\mathbf{y}, \mathbf{x}_i, \mathbf{x}_{-i}^{*}\right), \forall \mathbf{x}_i \in \Omega_i\left({\mathbf{y}, \mathbf{x}^{*}_{-i}}\right), \forall i \in \mathbf{F}   \right\}$ is a set of the followers' generalized Nash equilibrium when the leader's decision is $\mathbf{y}$,
$\mathbf{x}^{*}\left(\mathbf{y}\right) = \left(\mathbf{x}^{*}_{i}\left(\mathbf{y}\right)\right)_{i\in\mathbf{F}}$,
$\mathbf{x}^{*}_{-i}\left(\mathbf{y}\right) = \left(\mathbf{x}^{*}_{1}\left(\mathbf{y}\right), \cdots, \mathbf{x}^{*}_{i-1}\left(\mathbf{y}\right), \mathbf{x}^{*}_{i+1}\left(\mathbf{y}\right), \cdots, \mathbf{x}^{*}_{N}\left(\mathbf{y}\right)\right)$, $\mathbf{x}^{*}=\left(\mathbf{x}_{i}^{*}\right)_{i\in\mathbf{F}}$
, and $\mathbf{x}^{*}_{-i} = \left(\mathbf{x}^{*}_{1}, \cdots, \mathbf{x}^{*}_{i-1}, \mathbf{x}^{*}_{i+1}, \cdots, \mathbf{x}^{*}_{N}\right)$.
\end{definition}

Next, we define a variational Stackelberg equilibrium of the $1-N$ generalized Stackelberg game $\Gamma$ to prove the existence of a generalized Stackelberg equilibrium. A variational Stackelberg equilibrium $\left(\mathbf{y}^{*}, \left(\mathbf{x}_{i}^{*}\left(\mathbf{y}^{*}\right)\right)_{i\in\mathbf{F}}\right)$ is the joint decision where a leader optimize its decision $\mathbf{y}^{*}$ while the followers' joint decision is on the variational equilibrium of the subgame $\mathrm{G}\left(\mathbf{y}^{*}\right)$ in equation (\ref{eqn:subgame}).

\begin{definition}
\label{def3}
The joint decision $\left(\mathbf{y}^{*}, \mathbf{x}^{*}\left(\mathbf{y}^{*}\right)\right)\in\Omega_{\mathrm{L}}\times\prod_{i\in\mathbf{F}}{\Omega_i\left(\mathbf{y}^{*}, \mathbf{x}^{*}_{-i}\left(\mathbf{y}^{*}\right)\right)}$ is a variational Stackelberg equilibrium of $\Gamma=\left<\mathbf{F}, f_{\mathrm{L}}, \left(f_{i}\right)_{i\in\mathbf{F}}, \Omega_{\mathrm{L}}, \left(\Omega_{i}\right)_{i\in\mathbf{F}}\right>$ if 
\setlength{\arraycolsep}{0.0em}
\begin{eqnarray}
\sup_{\mathbf{x}^{*}\left(\mathbf{y}^{*}\right)\in \mathrm{VE}\left(\mathbf{y}^{*}\right)}f_{\mathrm{L}}\left(\mathbf{y}^{*}, \mathbf{x}^{*}\left(\mathbf{y}^{*}\right)\right)&\geq&\inf_{\mathbf{x}^{*}\left(\mathbf{y}\right)\in \mathrm{VE}\left(\mathbf{y}\right)}f_{\mathrm{L}}\left(\mathbf{y}, \mathbf{x}^{*}\left(\mathbf{y}\right)\right), \forall \mathbf{y} \in \Omega_{\mathrm{L}}
\label{eqn:def3}
\end{eqnarray}
where $\mathrm{VE}\left(\mathbf{y}\right)=\left\{ \mathbf{x}^{*} \in \prod_{i\in\mathbf{F}}\Omega_{i}(\mathbf{y},\mathbf{x}^{*}_{-i}) | \mathbf{D}\left(\mathbf{y}, \mathbf{x}^{*}\right)^{\mathrm{T}}\left(\mathbf{x}^{*}-\mathbf{x}\right)\geq 0, \forall \mathbf{x}\in\prod_{i\in\mathbf{F}}{\Omega_i\left(\mathbf{y}, \mathbf{x}_{-i}\right)}
\right\}$ is a set of the follower's variational equilibrium when the leader's decision is $\mathbf{y}$, $\mathbf{D}\left(\mathbf{y}, \mathbf{x}^{*}\right)=\left(\nabla_{\mathbf{x}_{i}}{f_i\left(\mathbf{y}, \mathbf{x}^{*}\right)}\right)_{i\in\mathbf{F}}$ is the gradient of the followers' objective, $\mathbf{x}^{*}\left(\mathbf{y}\right) = \left(\mathbf{x}^{*}_{i}\left(\mathbf{y}\right)\right)_{i\in\mathbf{F}}$,
$\mathbf{x}^{*}_{-i}\left(\mathbf{y}\right) = \left(\mathbf{x}^{*}_{1}\left(\mathbf{y}\right), \cdots, \mathbf{x}^{*}_{i-1}\left(\mathbf{y}\right), \mathbf{x}^{*}_{i+1}\left(\mathbf{y}\right), \cdots, \mathbf{x}^{*}_{N}\left(\mathbf{y}\right)\right)$, $\mathbf{x}^{*}=\left(\mathbf{x}_{i}^{*}\right)_{i\in\mathbf{F}}$,
$\mathbf{x}^{*}_{-i} = \left(\mathbf{x}^{*}_{1}, \cdots, \mathbf{x}^{*}_{i-1}, \mathbf{x}^{*}_{i+1}, \cdots, \mathbf{x}^{*}_{N}\right)$, $\mathbf{x}=\left(\mathbf{x}_{i}\right)_{i\in\mathbf{F}}$,
and $\mathbf{x}_{-i} = \left(\mathbf{x}_{1}, \cdots, \mathbf{x}_{i-1}, \mathbf{x}_{i+1}, \cdots, \mathbf{x}_{N}\right)$.

\end{definition}

The generalized Stackelberg equilibrium may not exist depending on the structures of the utility function and the strategy set. However, we can define the $1-N$ generalized Stackelberg game $\Gamma=\left<\mathbf{F}, f_{\mathrm{L}}, \left(f_{i}\right)_{i\in\mathbf{F}}, \Omega_{\mathrm{L}}, \left(\Omega_{i}\right)_{i\in\mathbf{F}}\right>$ that always have a generalized Stackelberg equilibrium by utilizing Theorems \ref{thm1} and \ref{thm2}.

\begin{theorem}
\label{thm3}
Let $\mathbf{D}\left(\cdot\right)=\left(\nabla_{\mathbf{x}_{i}}{f_i\left(\cdot\right)}\right)_{i\in\mathbf{F}}$ is the gradient of followers' objective.
If the following four conditions are satisfied, then the $1-N$ generalized Stackelberg game $\Gamma=\left<\mathbf{F}, f_{\mathrm{L}}, \left(f_{i}\right)_{i\in\mathbf{F}}, \Omega_{\mathrm{L}}, \left(\Omega_{i}\right)_{i\in\mathbf{F}}\right>$ has a variational Stackelberg equilibrium, and it is also a generalized Stackelberg equilibrium of $\Gamma$.

\begin{itemize}
    \item {C}
    1. $\Omega_i$ is closed and convex for {all} $i \in \mathbf{F}$
    \item {C}
    2. $f_i$ is concave $C^{1}$-function on $\Omega_i$ for {all} $i \in \mathbf{F}$
    \item {C}
    3. $\prod_{i\in\mathbf{F}}{\Omega_i}$ is jointly convex
    \item {C}
    4. $-\mathbf{D}\left(\cdot\right)$ is strongly monotone on $\prod_{i\in\mathbf{F}}{\Omega_i}$
\end{itemize}
\end{theorem}
The proof is provided in Appendix \ref{A.1}.

In addition, the uniqueness of the generalized Stackelberg equilibrium is guaranteed if some additional conditions are satisfied.

\begin{proposition}
\label{prop1}
Let $\Gamma=\left<\mathbf{F}, f_{\mathrm{L}}, \left(f_{i}\right)_{i\in\mathbf{F}}, \Omega_{\mathrm{L}}, \left(\Omega_{i}\right)_{i\in\mathbf{F}}\right>$ be a $1-N$ generalized Stackelberg game that satisfies the conditions of Theorem \ref{thm3}. Then, $\Gamma$ has the unique generalized Stackelberg equilibrium if the following three conditions are satisfied: (1) $f_{\mathrm{L}}\left(\mathbf{y}, \mathbf{x}\left(\mathbf{y}\right)\right)$ is strictly concave on $\mathbf{y}\in\Omega_{\mathrm{L}}$; (2) $\Omega_{\mathrm{L}}$ is closed and convex, and (3) the set of a generalized Nash equilibrium of $\mathrm{G}\left(\mathbf{y}\right)$ is unique for all $\mathbf{y}\in\Omega_{\mathrm{L}}$.
\end{proposition}
The proof is provided in Appendix \ref{A.2}.

Theorem \ref{thm3} states the conditions where a $1-N$ generalized Stackelberg game $\Gamma=\left<\mathbf{F}, f_{\mathrm{L}}, \left(f_{i}\right)_{i\in\mathbf{F}}, \Omega_{\mathrm{L}}, \left(\Omega_{i}\right)_{i\in\mathbf{F}}\right>$ always have a generalized Stackelberg equilibrium. In the remaining section, we assume that $\Gamma$ satisfies the conditions of Theorem \ref{thm3}.

\section{Computing Method}
To find an equilibrium in polynomial time, we develop a method to convert the $1-N$ generalized Stackelberg game $\Gamma=\left<\mathbf{F}, f_{\mathrm{L}}, \left(f_{i}\right)_{i\in\mathbf{F}}, \Omega_{\mathrm{L}}, \left(\Omega_{i}\right)_{i\in\mathbf{F}}\right>$ to the $1-1$ Stackelberg game $\hat{\Gamma}=\left<\left\{1\right\}, f_{\mathrm{L}}, f_{\mathrm{F}}, \Omega_{\mathrm{L}}, \Omega_{\mathrm{F}}\right>$ having the same solution (Section 5.1). We then propose the projected implicit gradient descent (PIGD) algorithm to find a Stackelberg equilibrium of the transformed $1-1$ Stackelberg game $\hat{\Gamma}$ (Section 5.2). Figure \ref{fig:overall} illustrates the procedure to compute a generalized Stackelberg equilibrium for the $1-N$ generalized Stackelberg game $\Gamma$. Finally, we prove that the proposed computing method finds equilibrium in polynomial time (Section 5.3).

\subsection{Transforming Method from the $1-N$ Generalized Stackelberg Game $\Gamma$ to the $1-1$ Stackelberg Game $\hat{\Gamma}$}
There is a generalized Stackelberg equilibrium of the $1-N$ generalized Stackelberg game $\Gamma$ while the four conditions of Theorem \ref{thm3} are satisfied. However, there have been no algorithms that can solve the generalized Stackelberg equilibrium of the $1-N$ generalized Stackelberg game in polynomial time. Therefore, to compute a generalized Stackelberg equilibrium in polynomial time, we transform the $1-N$ generalized Stackelberg game into a solvable problem. Specifically, we transform the $1-N$ generalized Stackelberg game $\Gamma$ into a $1-1$ Stackelberg game $\hat{\Gamma}$ whose Stackelberg equilibrium is identical to that of $\Gamma$.

First, we define the $1-1$ Stackelberg game $\hat{\Gamma}=\left<\left\{1\right\}, f_{\mathrm{L}}, f_{\mathrm{F}}, \Omega_{\mathrm{L}}, \Omega_{\mathrm{F}}\right>$ whose Stackelberg equilibrium is identical to that of $\Gamma=\left<\mathbf{F}, f_{\mathrm{L}}, \left(f_{i}\right)_{i\in\mathbf{F}}, \Omega_{\mathrm{L}}, \left(\Omega_{i}\right)_{i\in\mathbf{F}}\right>$ which can be formulated as follows:
\setlength{\arraycolsep}{0.0em}
\begin{eqnarray}
\mathbf{y}^{*} & = & \argmax_{\mathbf{y} \in \Omega_{\mathrm{L}}}{f_{\mathrm{L}}\left(\mathbf{y}, \mathbf{x}^{*}\left(\mathbf{y}\right) \right)}\nonumber\\
\mathbf{w}^{*} \left(\mathbf{y}\right) & = & \argmax_{\mathbf{w} \in \Omega_{\mathrm{F}}\left(\mathbf{y}\right)}{f_{\mathrm{F}}\left(\mathbf{y}, \mathbf{w}\right)}\nonumber\\
\mathrm{where} \quad \mathbf{w}&:=&\left(\mathbf{x}, \mathbf{z}, \boldsymbol{\lambda}, \boldsymbol{\mu}\right)\in\mathbb{R}^{n_{\mathrm{F}}}, \quad n_{\mathrm{F}}:=\sum\limits_{i\in\mathbf{F}}{\left(2n_i+p_i+q_i\right)}\nonumber\\
f_{\mathrm{F}}\left(\mathbf{y}, \mathbf{w}\right) &=& \left(\nabla_{\mathbf{x}_{i}} f_{i}\left(\mathbf{y}, \mathbf{x}\right)\right)_{i\in\mathbf{F}}^{\mathrm{T}}\left(\mathbf{x}-\mathbf{z}\right)\nonumber\\
\Omega_{\mathrm{L}} &=& \left\{\mathbf{y}\in\mathbb{R}^{n_{\mathrm{L}}} \middle|
\begin{matrix}
h^{j}_{\mathrm{L}}\left(\mathbf{y}\right) \leq 0, \forall j \in \left[ p_{\mathrm{L}} \right]\\
l^{k}_{\mathrm{L}}\left(\mathbf{y}\right) = 0, \forall k \in \left[ q_{\mathrm{L}} \right]
\end{matrix} \right\}\nonumber \\
\Omega_{\mathrm{F}}\left(\mathbf{y}\right) &=& \left\{\mathbf{w}\in\mathbb{R}^{n_{\mathrm{F}}} \middle|
\begin{matrix}
-\left(\nabla_{\mathbf{x}_{i}} f_{i}\left(\mathbf{y}, \mathbf{x}\right)\right)_{i\in\mathbf{F}}^{\mathrm{T}}+\sum\limits_{i\in\mathbf{F}}{\sum\limits_{j\in\left[p_{i}\right]}{\lambda_{i}^{j}\nabla_{\mathbf{z}}{h_{i}^{j}\left(\mathbf{y}, \mathbf{z}\right)}}}\\
+\sum\limits_{i\in\mathbf{F}}{\sum\limits_{k\in\left[q_i\right]}{\mu_{i}^{k}\nabla_{\mathbf{z}}{l_{i}^{k}\left(\mathbf{y}, \mathbf{z}\right)}}} = \mathbf{0}\\
h^{j}_{i}\left(\mathbf{y}, \mathbf{x}\right) \leq 0, \forall j\in\left[ p_i \right], \forall i\in\mathbf{F}\\
l^{k}_{i}\left(\mathbf{y}, \mathbf{x}\right) = 0, \forall k\in\left[ q_i \right], \forall i\in\mathbf{F}\\
\lambda_{i}^{j} h_{i}^{j}\left(\mathbf{y}, \mathbf{z}\right) = 0, \forall j\in\left[p_i\right], \forall i\in\mathbf{F}\\
h_{i}^{j}\left(\mathbf{y}, \mathbf{z}\right) \leq 0, \forall j\in\left[p_i\right], \forall i\in\mathbf{F}\\
l_{i}^{k}\left(\mathbf{y}, \mathbf{z}\right) = 0, \forall k\in\left[q_i\right], \forall i\in\mathbf{F}\\
\lambda_{i}^{j} \geq 0, \forall j\in\left[p_i\right], \forall i\in\mathbf{F}
\end{matrix}  \right\}\label{eqn:1-1 Stackelberg game}
\end{eqnarray}
where $f_{\mathrm{L}}$ be the objective function of the leader, $f_{\mathrm{F}}$ be the objective function of the follower, $\mathbf{y}$ is the leader's decision belonging to the strategy set $\Omega_{\mathrm{L}}$, and $\mathbf{w}$ is the follower's decision belonging to the strategy set $\Omega_{\mathrm{F}}\left(\mathbf{y}\right)$. To be specific, the follower's decision is composed of $\mathbf{x}=\left(\mathbf{x}_{i}\right)_{i\in\mathbf{F}}\in\mathbb{R}^{\sum\limits_{i\in\mathbf{F}}{n_i}}$, 
$\mathbf{z}=\left(\mathbf{z}_{i}\right)_{i\in\mathbf{F}}\in\mathbb{R}^{\sum\limits_{i\in\mathbf{F}}{n_i}}$, 
$\boldsymbol{\lambda}=\left(\lambda_{i}^{j}\right)_{i,j}\in\mathbb{R}^{\sum\limits_{i\in\mathbf{F}}{p_i}}$, and 
$\boldsymbol{\mu}=\left(\mu_{i}^{k}\right)_{i,k}\in\mathbb{R}^{\sum\limits_{i\in\mathbf{F}}{q_i}}$. Now, we prove that the Stackelberg equilibrium of $\hat{\Gamma}$ is equivalent to the generalized Stackelberg equilibrium of $\Gamma$.

\begin{theorem}
\label{thm4}
The set of the variational Stackelberg equilibrium of the $1-N$ generalized Stackelberg game $\Gamma=\left<\mathbf{F}, f_{\mathrm{L}}, \left(f_{i}\right)_{i\in\mathbf{F}}, \Omega_{\mathrm{L}}, \left(\Omega_{i}\right)_{i\in\mathbf{F}}\right>$ is equivalent to the set of the Stackelberg equilibrium of the $1-1$ Stackelberg game $\hat{\Gamma}=\left<\left\{1\right\}, f_{\mathrm{L}}, f_{\mathrm{F}}, \Omega_{\mathrm{L}}, \Omega_{\mathrm{F}}\right>$.
\end{theorem}

\proof{Proof of Theorem \ref{thm4}}
Let $\mathrm{G}\left(\mathbf{y}\right)=\left<\mathbf{F}, \left(f_i\right)_{i\in\mathbf{F}}, \left(\Omega_i\right)_{i\in\mathbf{F}}\right>$ be the $N$ followers' subgame of $\Gamma=\left<\mathbf{F}, f_{\mathrm{L}}, \left(f_{i}\right)_{i\in\mathbf{F}}, \Omega_{\mathrm{L}}, \left(\Omega_{i}\right)_{i\in\mathbf{F}}\right>$. By equation (\ref{eqn:def2}) of Definition \ref{def2}, the set of variational equilibrium of $\mathrm{G}\left(\mathbf{y}\right)$ is defined as
\setlength{\arraycolsep}{0.0em}
\begin{eqnarray}
\left\{\mathbf{x}\in\prod_{i\in\mathbf{F}}{\Omega_{i}\left(\mathbf{y},\mathbf{x}_{-i}\right)} \middle| \mathbf{D}\left(\mathbf{y}, \mathbf{x}\right)^{\mathrm{T}} \left(\mathbf{x} - \mathbf{z}\right)\geq 0, \forall \mathbf{z}\in\prod_{i\in\mathbf{F}}{\Omega_i\left(\mathbf{y}, \mathbf{z}_{-i}\right)}\right\}
\label{eqn:appendix_ve}
\end{eqnarray}
where $\mathbf{D}\left(\mathbf{y}, \mathbf{x}\right)=\left(\nabla_{\mathbf{x}_{i}}f_{i}\left(\mathbf{y}, \mathbf{x}\right)\right)_{i\in\mathbf{F}}$ is the gradient of followers' objective, $\mathbf{x}=\left(\mathbf{x}_{i}\right)_{i\in\mathbf{F}}$, $\mathbf{x}_{-i}=\left(\mathbf{x}_{1}, \cdots, \mathbf{x}_{i-1}, \mathbf{x}_{i+1}, \cdots, \mathbf{x}_{N}\right)$, $\mathbf{z}=\left(\mathbf{z}_{i}\right)_{i\in\mathbf{F}}$, and $\mathbf{z}_{-i}=\left(\mathbf{z}_{1}, \cdots, \mathbf{z}_{i-1}, \mathbf{z}_{i+1}, \cdots, \mathbf{z}_{N}\right)$. Then, the variational Stackelberg equilibrium of $\Gamma$ is the solution of the following decision-making problem by the definition of the variational equilibrium:
\setlength{\arraycolsep}{0.0em}
\begin{eqnarray}
&&\max_{\mathbf{y}\in\Omega_{\mathrm{L}}}{f_{\mathrm{L}}\left(\mathbf{y}, \mathbf{x}\right)}\nonumber\\
\mathrm{s.t.} &&\mathbf{D}\left(\mathbf{y}, \mathbf{x}\right)^{\mathrm{T}}\left(\mathbf{x}-\mathbf{z}\right)\geq 0, \forall \mathbf{z}\in\prod_{i\in\mathbf{F}}{\Omega_i\left(\mathbf{y}, \mathbf{z}_{-i}\right)}\nonumber\\&&\mathbf{x}\in\prod_{i\in\mathbf{F}}{\Omega_{i}\left(\mathbf{y},\mathbf{x}_{-i}\right)}\label{eqn:appendix3}
\end{eqnarray}

Let $\mathbf{z}^{*}\left(\mathbf{y}, \mathbf{x}\right)=\argmin_{\mathbf{z}\in\prod_{i\in\mathbf{F}}{\Omega_i\left(\mathbf{y}, \mathbf{z}_{-i}\right)}}{\mathbf{D}\left(\mathbf{y}, \mathbf{x}\right)^{\mathrm{T}}\left(\mathbf{x}-\mathbf{z}\right)}$. Obviously, the solution set of the problem (\ref{eqn:appendix3}) is equivalent to the solution set of the following decision-making problem:
\setlength{\arraycolsep}{0.0em}
\begin{eqnarray}
&&\max_{\mathbf{y}\in\Omega_{\mathrm{L}}}{f_{\mathrm{L}}\left(\mathbf{y}, \mathbf{x}\right)}\nonumber\\
\mathrm{s.t.} &&\mathbf{D}\left(\mathbf{y}, \mathbf{x}\right)^{\mathrm{T}}\left(\mathbf{x}-\mathbf{z}^{*}\left(\mathbf{y}, \mathbf{x}\right)\right)\geq 0, \mathbf{x}\in\prod_{i\in\mathbf{F}}{\Omega_{i}\left(\mathbf{y},\mathbf{x}_{-i}\right)}
\label{eqn:appendix4-1}
\end{eqnarray}

There is the unique variational equilibrium of $\mathrm{G}\left(\mathbf{y}\right)$ by Theorem \ref{thm1} since three conditions are satisfied: (1) $\Omega_i$ is closed and convex for $i\in\mathbf{F}$ by the first condition of Theorem \ref{thm3}; (2) $f_i$ is continuous on $\Omega_i$ for $i\in\mathbf{F}$ by the second condition of Theorem \ref{thm3}; and (3) $\left(-\nabla_{\mathbf{x}_{i}}f_{i}\left(\mathbf{y}, \mathbf{x}\right)\right)_{i\in\mathbf{F}}$ is strongly monotone on $\prod_{i\in\mathbf{F}}\Omega_i$ by the fourth condition of Theorem \ref{thm3}. It means that the variational equilibrium $\mathbf{x}^{*}\left(\mathbf{y}\right)$ of $\mathrm{G}\left(\mathbf{y}\right)$ is unique and always exists for all the leader's decisions $\mathbf{y}\in\Omega_{\mathrm{L}}$.

Let $\hat{\mathbf{x}}\left(\mathbf{y}\right)=\argmax_{\mathbf{x}\in\prod_{i\in\mathbf{F}}{\Omega_{i}\left(\mathbf{y}, \mathbf{x}_{-i}\right)}}{\mathbf{D}\left(\mathbf{y}, \mathbf{x}\right)^{\mathrm{T}}\left(\mathbf{x} - \mathbf{z}^{*}\left(\mathbf{y}, \mathbf{x}\right)\right)}$. Then, the following equation holds by the definition of $\hat{\mathbf{x}}\left(\mathbf{y}\right)$.
\setlength{\arraycolsep}{0.0em}
\begin{eqnarray}
\mathbf{D}\left(\mathbf{y}, \hat{\mathbf{x}}\left(\mathbf{y}\right)\right)^{\mathrm{T}}\left(\hat{\mathbf{x}}\left(\mathbf{y}\right) - \mathbf{z}^{*}\left(\mathbf{y}, \hat{\mathbf{x}}\left(\mathbf{y}\right)\right)\right) \geq \mathbf{D}\left(\mathbf{y}, \mathbf{x}^{*}\left(\mathbf{y}\right)\right)^{\mathrm{T}}\left(\mathbf{x}^{*}\left(\mathbf{y}\right) - \mathbf{z}^{*}\left(\mathbf{y}, \mathbf{x}^{*}\left(\mathbf{y}\right)\right)\right)
\label{eqn:sub1}
\end{eqnarray}
By the definition of $\mathbf{x}^{*}\left(\mathbf{y}\right)$, equation (\ref{eqn:appendix_ve}) is not empty.
\setlength{\arraycolsep}{0.0em}
\begin{eqnarray}
\mathbf{D}\left(\mathbf{y}, \mathbf{x}^{*}\left(\mathbf{y}\right) \right)^{\mathrm{T}}\left(\mathbf{x}^{*}\left(\mathbf{y}\right) - \mathbf{z}^{*}\left(\mathbf{y}, \mathbf{x}^{*}\left(\mathbf{y}\right)\right)\right) \geq 0
\label{eqn:sub2}
\end{eqnarray}
When we substitute equation (\ref{eqn:sub2}) into equation (\ref{eqn:sub1}), we obtain $\mathbf{D}\left(\mathbf{y}, \hat{\mathbf{x}}\left(\mathbf{y}\right)\right)^{\mathrm{T}}\left(\hat{\mathbf{x}}\left(\mathbf{y}\right) - \mathbf{z}^{*}\left(\mathbf{y}, \hat{\mathbf{x}}\left(\mathbf{y}\right)\right)\right) \geq 0$, that is, $\hat{\mathbf{x}}\left(\mathbf{y}\right)$ is also a variational equilibrium of $\mathrm{G}\left(\mathbf{y}\right)$. Since $\mathrm{G}\left(\mathbf{y}\right)$ has the unique variational equilibrium for all $\mathbf{y}\in\Omega_{\mathrm{L}}$, $\mathbf{x}^{*}\left(\mathbf{y}\right)$ is equal to $\hat{\mathbf{x}}\left(\mathbf{y}\right)$. Therefore, the solution set of the problem (\ref{eqn:appendix4-1}) is equivalent to the solution set of the following optimization problem:
\setlength{\arraycolsep}{0.0em}
\begin{eqnarray}
&&\max_{\mathbf{y}\in\Omega_{\mathrm{L}}}{f_{\mathrm{L}}\left(\mathbf{y}, \mathbf{x}^{*}\left(\mathbf{y}\right)\right)}\nonumber\\
\mathrm{s.t.} && \mathbf{x}^{*}\left(\mathbf{y}\right)=\argmax_{\mathbf{x}\in\prod_{i\in\mathbf{F}}{\Omega_{i}\left(\mathbf{y}, \mathbf{x}_{-i}\right)}}{\mathbf{D}\left(\mathbf{y}, \mathbf{x}\right)^{\mathrm{T}}\left(\mathbf{x}-\mathbf{z}^{*}\left(\mathbf{y}, \mathbf{x}\right)\right)}\nonumber\\
\mathrm{s.t.} && \mathbf{z}^{*}\left(\mathbf{y}, \mathbf{x}\right) = \argmin_{\mathbf{z}\in\prod_{i\in\mathbf{F}}{\Omega_i\left(\mathbf{y}, \mathbf{z}_{-i}\right)}}{\mathbf{D}\left(\mathbf{y}, \mathbf{x}\right)^{\mathrm{T}}\left(\mathbf{x}-\mathbf{z}\right)}
\label{eqn:appendix5}
\end{eqnarray}

Finally, we transform the problem (\ref{eqn:appendix5}) into the $1-1$ Stackelberg game $\hat{\Gamma}=\left<\left\{1\right\}, f_{\mathrm{L}}, f_{\mathrm{F}}, \Omega_{\mathrm{L}}, \Omega_{\mathrm{F}}\right>$ by using the KKT conditions of $\mathbf{z}^{*}\left(\mathbf{y}, \mathbf{x}\right) = \argmin_{\mathbf{z}\in\prod_{i\in\mathbf{F}}{\Omega_i\left(\mathbf{y}, \mathbf{z}_{-i}\right)}}{\mathbf{D}\left(\mathbf{y}, \mathbf{x}\right)^{\mathrm{T}}\left(\mathbf{x}-\mathbf{z}\right)}$. 
\setlength{\arraycolsep}{0.0em}
\begin{eqnarray}
&&\max_{\mathbf{y}\in\Omega_{\mathrm{L}}}{f_{\mathrm{L}}\left(\mathbf{y}, \mathbf{x}^{*}\left(\mathbf{y}\right)\right)}\nonumber\\
\mathrm{s.t.} && \left(\mathbf{x}^{*}\left(\mathbf{y}\right), \mathbf{z}^{*}\left(\mathbf{y}\right), \boldsymbol{\lambda}\left(\mathbf{y}\right), \boldsymbol{\mu}\left(\mathbf{y}\right) \right)=\argmax_{\left(\mathbf{x}, \mathbf{z}, \boldsymbol{\lambda}, \boldsymbol{\mu}\right) \in \prod_{i\in\mathbf{F}}{\Omega_{i}\left(\mathbf{y}, \mathbf{x}_{-i}\right)}\cap\mathbf{S}}{\mathbf{D}\left(\mathbf{y}, \mathbf{x}\right)^{\mathrm{T}}\left(\mathbf{x}-\mathbf{z}\right)}
\label{eqn:appendix6}
\end{eqnarray}
where $\boldsymbol{\lambda}=\left(\lambda_{i}^{j}\right)_{i, j}$, $\boldsymbol{\mu}=\left(\mu_{i}^{k}\right)_{i, k}$, and $\mathbf{S}$ is the set of the KKT conditions of $\mathbf{z}^{*}\left(\mathbf{y}, \mathbf{x}\right) = \argmin_{\mathbf{z}\in\prod_{i\in\mathbf{F}}{\Omega_i\left(\mathbf{y}, \mathbf{z}_{-i}\right)}}{\mathbf{D}\left(\mathbf{y}, \mathbf{x}\right)^{\mathrm{T}}\left(\mathbf{x}-\mathbf{z}\right)}$. To be specifically, $\mathbf{S}$ is defined as follows:
\begin{equation}
\mathbf{S}=
\left\{
\mathbf{x}, \mathbf{z}, \boldsymbol{\lambda}, \boldsymbol{\mu}
\middle|
\begin{matrix}
-\mathbf{D}\left(\mathbf{y}, \mathbf{x}\right)^{\mathrm{T}}+\sum\limits_{i\in\mathbf{F}}{\sum\limits_{j\in\left[p_{i}\right]}{\lambda_{i}^{j}\nabla_{\mathbf{z}}{h_{i}^{j}\left(\mathbf{y}, \mathbf{z}\right)}}}
+ \sum\limits_{i\in\mathbf{F}}{\sum\limits_{k\in\left[q_i\right]}{\mu_{i}^{k}\nabla_{\mathbf{z}}{l_{i}^{k}\left(\mathbf{y}, \mathbf{z}\right)}}} = \mathbf{0}
\\
  \lambda_{i}^{j} h_{i}^{j}\left(\mathbf{y}, \mathbf{z}\right) = 0, \forall j\in\left[p_i\right], \forall i\in\mathbf{F}
\\
 h_{i}^{j}\left(\mathbf{y}, \mathbf{z}\right) \leq 0, \forall j\in\left[p_i\right], l_{i}^{k}\left(\mathbf{y}, \mathbf{z}\right) = 0, \forall k\in\left[q_i\right], \forall i\in\mathbf{F}
\\
 \lambda_{i}^{j} \geq 0, \forall j\in\left[p_i\right], \forall i\in\mathbf{F}
\end{matrix}
\right\}\label{eqn:S}
\end{equation}

Because $f_{\mathrm{F}}=\mathbf{D}\left(\mathbf{y}, \mathbf{x}\right)^{\mathrm{T}}\left(\mathbf{x}-\mathbf{z}\right)$ and $\Omega_{\mathrm{F}}$ is equivalent to $\prod_{i\in\mathbf{F}}\Omega_{i}\left(\mathbf{y}, \mathbf{x}_{-i}\right)\cap\mathbf{S}$, the solution set of the problem (\ref{eqn:appendix6}) is equivalent to the Stackelberg equilibrium of the $1-1$ Stackelberg game $\hat{\Gamma}=\left<\left\{1\right\}, f_{\mathrm{L}}, f_{\mathrm{F}}, \Omega_{\mathrm{L}}, \Omega_{\mathrm{F}}\right>$. It means that the set of the Stackelberg equilibrium of $\hat{\Gamma}$ is equivalent to the set of the variational Stackelberg equilibrium of $\Gamma$. \Halmos
\endproof

Thus, we can compute a variational Stackelberg equilibrium of the $1-N$ generalized Stackelberg game $\Gamma=\left<\mathbf{F}, f_{\mathrm{L}}, \left(f_i\right)_{i\in\mathbf{F}}, \Omega_{\mathrm{L}}, \left(\Omega_i\right)_{i\in\mathbf{F}}\right>$ by computing a Stackelberg equilibrium of the $1-1$ Stackelberg game $\hat{\Gamma}=\left<\left\{1\right\}, f_{\mathrm{L}}, f_{\mathrm{F}}, \Omega_{\mathrm{L}}, \Omega_{\mathrm{F}}\right>$. Then, the computed variational Stackelberg equilibrium is a generalized Stackelberg equilibrium by Theorem \ref{thm3}. Next, we discuss how to compute the Stackelberg equilibrium of the $1-1$ Stackelberg game $\hat{\Gamma}$.

\subsection{Computing Method for the $1-1$ Stackelberg Game $\hat{\Gamma}$}
\label{subsec:computation}
To formally model the decision-making procedure of the $1-1$ Stackelberg game $\hat{\Gamma}=\left<\left\{1\right\}, f_{\mathrm{L}}, f_{\mathrm{F}}, \Omega_{\mathrm{L}}, \Omega_{\mathrm{F}}\right>$, let $f_{\mathrm{L}}$ be the objective function of the leader, and $\mathbf{y}$ is the leader's decision belonging to the strategy set $\Omega_{\mathrm{L}}$. A leader solves the following decision-making problem:
\begin{eqnarray}
\max_{\mathbf{y} \in \Omega_{\mathrm{L}}}{f_{\mathrm{L}}\left(\mathbf{y}, \mathbf{x}^{*}\left(\mathbf{y}\right)\right)}
\label{eqn:A}
\end{eqnarray}
Once a leader makes a decision, a follower optimizes its objective function $f_{\mathrm{F}}$ as follows:
\begin{eqnarray}
    \begin{aligned}
    \max_{\mathbf{w} \in \Omega_{\mathrm{F}}\left(\mathbf{y}\right)}{f_{\mathrm{F}}\left(\mathbf{y}, \mathbf{w}\right)}
    \label{eqn:B}
    \end{aligned}
\end{eqnarray}
where $\mathbf{w} = \left(\mathbf{x}, \mathbf{z}, \boldsymbol{\lambda}, \boldsymbol{\mu}\right)$ is the follower's decision variable belonging to the strategy set $\Omega_{\mathrm{F}}\left(\mathbf{y}\right)$.

We apply the projected implicit gradient descent (PIGD) algorithm to solve the $1-1$ Stackelberg game $\hat{\Gamma}$. The leader iteratively updates its decision to increase the objective until reaching the stationary point during which the follower (the fictitious follower that actually represents the $N$ followers in the original $1-N$ generalized Stackelberg game) also iteratively updates its decision to maximize its objective. In solving this problem, it is essential to track how the leader's decision affects the solution of the lower-level problem and tracks back its influence for optimizing the leader's decision. Algorithm \ref{alg:1} summarizes the overall process of computing the Stackelberg equilibrium of the $1-1$ Stackelberg game using the PIGD algorithm.

\begin{algorithm}
\caption{PIGD Algorithm for $1-1$ Stackelberg Game $\hat{\Gamma}$}
    \begin{algorithmic}[1]
        \Require{Feasible leader's decision $\mathbf{y}^{(0)}$, updating coefficient $\rho>0$} and  stopping criterion $\epsilon>0$.
        \Repeat \ for $t=0,1,2,\dots, t_{\mathrm{max}}$
            \State \textbf{Compute the lower-level solution} $\mathbf{w}^{(t)}=\left(\mathbf{x}^{(t)},\mathbf{z}^{(t)},\boldsymbol{\lambda}^{(t)},\boldsymbol{\mu}^{(t)}\right)$ given the leader's decision $\mathbf{y}^{(t)}$
            \State \textbf{Compute the implicit gradient} of the leader's objective with respect to the leader's decision based on the chain rule
                \begin{eqnarray}
                    \frac{d f_{\mathrm{L}}\left(\mathbf{y}^{(t)},\mathbf{w}^{(t)}\right)}{d \mathbf{y}}=\frac{\partial f_{\mathrm{L}}\left(\mathbf{y}^{(t)},\mathbf{w}^{(t)}\right)}{\partial \mathbf{y}} + \frac{\partial f_{\mathrm{L}}\left(\mathbf{y}^{(t)},\mathbf{w}^{(t)}\right)}{\partial \mathbf{w}}\cdot\frac{d \mathbf{w}^{\left(t\right)}\left(\mathbf{y}^{(t)}\right)}{d \mathbf{y}}
                    \label{eqn:grad}
                \end{eqnarray}
            \State \textbf{Update the leader decision} using the projected gradient descent
            \begin{eqnarray}
            \mathbf{y}^{(t+1)}=\mathrm{proj}_{\Omega_{\mathrm{L}}}\left(\mathbf{y}^{(t)}+\rho \frac{d f_{\mathrm{L}}\left(\mathbf{y}^{(t)},\mathbf{w}^{(t)}\right)}{d \mathbf{y}}\right)
            \label{eqn:PG}
            \end{eqnarray}
        \Until{$\left\|\mathbf{y}^{(t+1)}-\mathbf{y}^{(t)}\right\|<\epsilon$}
    \end{algorithmic}
\label{alg:1}
\end{algorithm}

Algorithm \ref{alg:1} is an iterative algorithm composed of the updating procedure of the leader's decision and the solution computing procedure of the follower. To apply Algorithm \ref{alg:1}, the solution $\mathbf{w}^{(t)}$ of the lower-level problem and the gradient of the leader's objective with respect to the leader's decision $\frac{d}{d \mathbf{y}} f_{\mathrm{L}}\left(\mathbf{y}^{(t)},\mathbf{w}^{(t)}\right)$ should be computed per every iteration.

\subsubsection{Computing $\mathbf{w}^{(t)}$ from given $\mathbf{y}^{(t)}$.}

Given the leader's decision $\mathbf{y}^{(t)}$, the lower-level solution $\mathbf{w}^{(t)}=\left(\mathbf{x}^{(t)},\mathbf{z}^{(t)},\boldsymbol{\lambda}^{(t)},\boldsymbol{\mu}^{(t)}\right)$ is computed by applying the variational equilibrium concept. The solution can be simply induced from the variational equilibrium of the followers' subgame $\mathrm{G}\left(\mathbf{y}\right) = \left<\mathbf{F}, \left( f_{i} \right)_{i \in \mathbf{F}}, \left(\Omega_{i}\right)_{i\in\mathbf{F}}\right>$ of the original $1-N$ generalized Stackelberg game $\Gamma=\left<\mathbf{F}, f_{\mathrm{L}}, \left(f_{i}\right)_{i\in\mathbf{F}}, \Omega_{\mathrm{L}}, \left(\Omega_{i}\right)_{i\in\mathbf{F}}\right>$ because Theorem \ref{thm4} proves that $\mathbf{x}^{(t)}$, which is a component of the solution $\mathbf{w}^{(t)}$, is equivalent to the variational equilibrium of $\mathrm{G}\left(\mathbf{y}\right)$.

To compute the variational equilibrium, various algorithms such as projected gradient descent \citep{malitsky2015projected} and NI-function type method \citep{facchinei2007generalized} can be employed. In this study, we compute the variational equilibrium of $\mathrm{G}\left(\mathbf{y}\right)$ by applying the projected gradient descent
\begin{eqnarray}
    \begin{aligned}
    \mathbf{x}^{(k+1)}= \mathrm{proj}_{\mathbf{C}^{\left(k\right)}}\left(\mathbf{x}^{(k)}+\rho\mathbf{D}\left(\mathbf{y}^{(t)}, \mathbf{x}^{(k)}\right)\right)
    \label{eqn:E}
    \end{aligned}
\end{eqnarray}
where $\mathbf{C}^{\left(k\right)}=\prod_{i\in\mathbf{F}}{\Omega_{i}\left(\mathbf{y}^{(t)}, \mathbf{x}^{(k)}_{-i}\right)}$ is the feasible reason, and $\rho$ is a positive step size. $\mathbf{z}^{(t)}$ is identical to $\mathbf{x}^{(t)}$ because $\mathbf{z}^{(t)}=\argmin\limits_{\mathbf{z}\in\prod_{i\in\mathbf{F}}{\Omega_i\left(\mathbf{y}^{(t)}, \mathbf{z}_{-i}\right)}}\mathbf{D}\left(\mathbf{y}^{(t)}, \mathbf{x}^{(k)}\right)^{\mathrm{T}}\left(\mathbf{x}^{(t)}-\mathbf{z}\right)$ and the objective is non-negative by the definition of the variational equilibrium. After computing $\mathbf{x}^{(t)}$ and $\mathbf{z}^{(t)}$, we can compute the remaining solutions, $\boldsymbol{\lambda}^{(t)}$ and $\boldsymbol{\mu}^{(t)}$, since they are feasible in the set $\mathbf{S}$ described in equation (\ref{eqn:S}).

\subsubsection{Computing the gradient of the leader's objective.}
The gradient of the leader's objective $f_{\mathrm{L}}\left(\mathbf{y}^{(t)},\mathbf{w}^{(t)}\right)$ with respect to its decision $\mathbf{y}$ given the lower-level solution $\mathbf{w}^{(t)}$ can be computed using the chain rule, as shown in equation (\ref{eqn:grad}). In the chain rule, $\frac{\partial}{\partial\mathbf{y}} f_{\mathrm{L}}\left(\mathbf{y}^{(t)},\mathbf{w}^{(t)}\right)$ and $\frac{\partial }{\partial \mathbf{w}}f_{\mathrm{L}}\left(\mathbf{y}^{(t)},\mathbf{w}^{(t)}\right)$ can be computed explicitly from the given leader's objective function $f_{\mathrm{L}}$. The challenging part is how to compute $\frac{d}{d \mathbf{y}} \mathbf{w}^{\left(t\right)}\left(\mathbf{y}^{(t)}\right)$, that is the gradient of the follower's decision $\mathbf{w}^{\left(t\right)}\left(\mathbf{y}\right)$ with respect to the leader's decision $\mathbf{y}$. In most cases, we cannot differentiate the follower's decision $\mathbf{w}^{\left(t\right)}\left(\mathbf{y}\right)$ with respect to the leader's decision $\mathbf{y}$ since the explicit formula of the $\mathbf{w}^{\left(t\right)}\left(\mathbf{y}\right)$ in terms of $\mathbf{y}$ is unknown. Thus, we apply the implicit differentiation techniques to compute $\frac{d}{d \mathbf{y}} \mathbf{w}^{\left(t\right)}\left(\mathbf{y}^{(t)}\right)$ \citep{gould2021deep}. 

When computing the gradient, handling inequality constraints in the follower's optimization problem is often complicated and intractable. To resolve this difficulty, \cite{gould2021deep} propose a simplified method for computing the implicit gradient while considering only the active inequality constraints (i.e., the inequality constraints whose value becomes zero given the current solution). Note that this choice can be justified in that a small step size $\rho$ results in small changes of the leader's and followers' decisions. That is, the inactive inequality constraints of the follower's strategy set $\Omega_{\mathrm{F}}$ will remain inactive after one iteration of updating the leader's decision with the gradient descent. In this sense, the inactive inequality constraints can be neglected while computing the gradient of the followers' decision with respect to the leader's decision in the $1-1$ Stackelberg game $\hat{\Gamma}$. The activity of the inequality constraints of $\Omega_{\mathrm{F}}\left(\mathbf{y}\right)$ is directly determined by $\boldsymbol{\lambda}^{\left(t\right)}$.

Thus, we rewrite the follower's strategy set of $\hat{\Gamma}$ at step $t$, $\Omega_{\mathrm{F}}^{\left(t\right)}\left(\mathbf{y}^{\left(t\right)}\right)$, consisting of the active inequality constraints and the equality constraints of the strategy set $\Omega_{\mathrm{F}}\left(\mathbf{y}\right)$ in the problem (\ref{eqn:1-1 Stackelberg game}):
\begin{eqnarray}
\Omega_{\mathrm{F}}^{\left(t\right)}\left(\mathbf{y}^{\left(t\right)}\right)=\left\{\mathbf{w}\in\mathbb{R}^{n_{\mathrm{w}}}\middle| l_{k}^{\left(t\right)}\left(\mathbf{y}^{\left(t\right)}, \mathbf{w}\right)=0, \forall k \in \left[q_{\mathrm{F}}^{\left(t\right)}\right] \right\}
\end{eqnarray}
where $l_{k}^{\left(t\right)}$ represents the $k$-th equality constraint at step $t$. Then, we can express the Lagrangian of the lower-level problem as $\mathcal{L}\left(\mathbf{y}^{\left(t\right)},\mathbf{w}^{\left(t\right)},\boldsymbol{\lambda}^{\left(t\right)}\right) = f_{\mathrm{F}}\left(\mathbf{y}^{\left(t\right)}, \mathbf{w}^{\left(t\right)}\right) + \sum\limits_{k \in \left[q_{\mathrm{F}}^{\left(t\right)}\right]}{\lambda}_{k}^{\left(t\right)} l_{k}^{\left(t\right)}\left(\mathbf{y}^{\left(t\right)}, \mathbf{w}^{\left(t\right)}\right)$ where $\boldsymbol{\lambda}^{\left(t\right)} = \left(\lambda_{k}^{\left(t\right)}\right)_{k \in \left[q_{\mathrm{F}}^{\left(t\right)}\right]}$.

Let's assume that $M_{\mathrm{F}}^{\left(t\right)} = \left(\nabla_{\mathbf{w}}l_{k}^{\left(t\right)}\left(\mathbf{y}^{\left(t\right)}, \mathbf{w}^{\left(t\right)}\right)\right)_{k \in \left[q_{\mathrm{F}}^{\left(t\right)}\right]}$ is full rank. Then, the lower-level solution $\mathbf{w}^{\left(t\right)}$ has a corresponding Lagrange multiplier $\boldsymbol{\lambda}^{\left(t\right)}$ such that $\left(\mathbf{w}^{\left(t\right)}, \boldsymbol{\lambda}^{\left(t\right)}\right)$ is a stationary point of the Lagrangian. In other words, at the stationary point, the gradient of the Lagrangian with respect to the follower's decision variable becomes a zero vector
\begin{eqnarray}
\label{eqn:lag_grad}
\nabla_{\mathbf{w}}\mathcal{L}\left(\mathbf{y}^{\left(t\right)},\mathbf{w}^{\left(t\right)},\boldsymbol{\lambda}^{\left(t\right)}\right) &=& \nabla_{\mathbf{w}}f_{\mathrm{F}}\left(\mathbf{y}^{\left(t\right)},\mathbf{w}^{\left(t\right)}\right) + {\boldsymbol{\lambda}^{\left(t\right)}}^{\mathrm{T}} M_{\mathrm{F}}^{\left(t\right)} = \mathbf{0}\nonumber\\
\nabla_{\boldsymbol{\lambda}}\mathcal{L}\left(\mathbf{y}^{\left(t\right)},\mathbf{w}^{\left(t\right)},\boldsymbol{\lambda}^{\left(t\right)}\right) &=& \left(l_{k}^{\left(t\right)}\left(\mathbf{y}^{\left(t\right)}, \mathbf{w}^{\left(t\right)}\right)\right)_{k\in\left[q_{\mathrm{F}}^{\left(t\right)}\right]}^{\mathrm{T}}=\mathbf{0}
\end{eqnarray}

We can compute the gradient of the follower's decision with respect to the leader's decision by differentiating the gradient of the Lagrangian in equation (\ref{eqn:lag_grad}) with respect to the leader's decision variable as:
\begin{eqnarray}
\frac{d \mathbf{w}^{\left(t\right)}\left(\mathbf{y}^{\left(t\right)}\right)}{d \mathbf{y}} &=& {M_{\mathrm{FF}}^{\left(t\right)}}^{-1} {M^{\left(t\right)}_{\mathrm{F}}}^{\mathrm{T}} \left({M^{\left(t\right)}_{\mathrm{F}}} {M^{\left(t\right)}_{\mathrm{FF}}}^{-1} {M^{\left(t\right)}_{\mathrm{F}}}^{\mathrm{T}}\right)^{-1} \nonumber\\
&\times& \left({M^{\left(t\right)}_{\mathrm{F}}} {M^{\left(t\right)}_{\mathrm{FF}}}^{-1} {M^{\left(t\right)}_{\mathrm{LF}}} - {M^{\left(t\right)}_{\mathrm{L}}}\right)-{M^{\left(t\right)}_{\mathrm{FF}}}^{-1} {M^{\left(t\right)}_{\mathrm{LF}}}\label{eqn:gradient of follower w.r.t. leader}
\end{eqnarray}
where

\begin{eqnarray}
M_{\mathrm{F}}^{\left(t\right)} &=& \left(\nabla_{\mathbf{w}}l_{k}^{\left(t\right)}\left(\mathbf{y}^{\left(t\right)}, \mathbf{w}^{\left(t\right)}\right)\right)_{k \in \left[q_{\mathrm{F}}^{\left(t\right)}\right]}\nonumber\\
M_{\mathrm{LF}}^{\left(t\right)} &=& -\nabla^{2}_{\mathbf{y}\mathbf{w}}f_{\mathrm{F}}\left(\mathbf{y}^{\left(t\right)},\mathbf{w}^{\left(t\right)}\right)-\sum_{k\in\left[q_{\mathrm{F}}^{\left(t\right)}\right]}\lambda_{k}^{\left(t\right)}\nabla^{2}_{\mathbf{y}\mathbf{w}}l_{k}^{\left(t\right)}\left(\mathbf{y}^{\left(t\right)}, \mathbf{w}^{\left(t\right)}\right)\nonumber\\
M_{\mathrm{L}}^{\left(t\right)} &=& \left(\nabla_{\mathbf{y}}l_{k}^{\left(t\right)}\left(\mathbf{y}^{\left(t\right)}, \mathbf{w}^{\left(t\right)}\right)\right)_{k \in \left[q_{\mathrm{F}}^{\left(t\right)}\right]}\nonumber\\
M_{\mathrm{FF}}^{\left(t\right)} &=& -\nabla^{2}_{\mathbf{w}\mathbf{w}}f_{\mathrm{F}}\left(\mathbf{y}^{\left(t\right)}, \mathbf{w}^{\left(t\right)}\right) - \sum_{k\in\left[q_{\mathrm{F}}^{\left(t\right)}\right]}\lambda_{k}^{\left(t\right)}\nabla^{2}_{\mathbf{w}\mathbf{w}}l_{k}^{\left(t\right)}\left(\mathbf{y}^{\left(t\right)}, \mathbf{w}^{\left(t\right)}\right)
\label{eqn:FF}
\end{eqnarray}
and $\boldsymbol{\lambda}=-\left(M^{\left(t\right)}_{\mathrm{F}} {M^{\left(t\right)}_{\mathrm{F}}}^{\mathrm{T}}\right)^{-1}M^{\left(t\right)}_{\mathrm{F}}\left(\nabla_{\mathbf{w}}f_{\mathrm{F}}\left(\mathbf{y}^{\left(t\right)},\mathbf{w}^{\left(t\right)}\right)\right)^{\mathrm{T}}$ by Proposition 4.6 of \citep{gould2021deep}. 

To compute the gradient of the follower's decision with respect to the leader's decision using equation (\ref{eqn:gradient of follower w.r.t. leader}), $M_{\mathrm{FF}}^{\left(t\right)}$ should be non-singular. Furthermore, the existence and uniqueness of the Lagrange multipliers are guaranteed only when $M_{\mathrm{F}}^{\left(t\right)}$ is a full rank matrix. However, these conditions do not hold for the general framework. Thus, we use pseudo-inverse to approximate the gradient of the follower's decision with respect to the leader's decision. Once it is computed, the gradient of the leader's objective with respect to the leader's decision $\frac{d f_{\mathrm{L}}\left(\mathbf{y},\mathbf{w}\right)}{d \mathbf{y}}$ is computed by equation (\ref{eqn:grad}).

\subsubsection{Updating the leader's decision.}
Once $\frac{d f_{\mathrm{L}}\left(\mathbf{y}^{(t)},\mathbf{w}^{(t)}\right)}{d \mathbf{y}}$ is computed, we can then update the leader's decision using equation (\ref{eqn:PG}).

In summary, we provide a PIGD algorithm that can be applied to the transformed $1-1$ Stackelberg game $\hat{\Gamma}=\left<\left\{1\right\}, f_{\mathrm{L}}, f_{\mathrm{F}}, \Omega_{\mathrm{L}}, \Omega_{\mathrm{F}}\right>$. First, we find the lower-level solution of $\hat{\Gamma}$ in each iteration by computing the variational equilibrium of the $N$ followers' subgame $\mathrm{G}\left(\mathbf{y}\right) = \left<\mathbf{F}, \left( f_{i} \right)_{i \in \mathbf{F}}, \left(\Omega_{i}\right)_{i\in\mathbf{F}}\right>$ of the $1-N$ generalized Stackelberg game $\Gamma=\left<\mathbf{F}, f_{\mathrm{L}}, \left(f_i\right)_{i\in\mathbf{F}}, \Omega_{\mathrm{L}}, \left(\Omega_i\right)_{i\in\mathbf{F}}\right>$ using the well-known projected gradient descent algorithm \citep{malitsky2015projected}. After obtaining the lower-level solution, we approximate the gradient of the leader's objective with respect to the leader's decision using the implicit differentiation techniques \citep{gould2021deep}.

\subsection{Time Complexity of Computing Method}

In the previous subsection, we develop a method to transform the $1-N$ generalized Stackelberg game $\Gamma=\left<\mathbf{F}, f_{\mathrm{L}}, \left(f_i\right)_{i\in\mathbf{F}}, \Omega_{\mathrm{L}}, \left(\Omega_i\right)_{i\in\mathbf{F}}\right>$ into a $1-1$ Stackelberg game $\hat{\Gamma}=\left<\left\{1\right\}, f_{\mathrm{L}}, f_{\mathrm{F}}, \Omega_{\mathrm{L}}, \Omega_{\mathrm{F}}\right>$ (Section 5.1) and propose a PIGD algorithm to find a Stackelberg equilibrium of $\hat{\Gamma}$ (Section 5.2). Now, we prove that our PIGD algorithm finds the Stackelberg equilibrium of $\hat{\Gamma}$ in polynomial time.

\begin{theorem}
\label{thm5}
Let $\Gamma=\left<\mathbf{F}, f_{\mathrm{L}}, \left(f_i\right)_{i\in\mathbf{F}}, \Omega_{\mathrm{L}}, \left(\Omega_i\right)_{i\in\mathbf{F}}\right>$ be a $1-N$ generalized Stackelberg game where the strategy set is defined as
\begin{eqnarray}
\Omega_{\mathrm{L}} &=& \left\{\mathbf{y}\in\mathbb{R}^{n_{\mathrm{L}}} \middle|
\begin{matrix}
h^{j}_{\mathrm{L}}\left(\mathbf{y}\right) \leq 0, \forall j \in \left[ p_{\mathrm{L}} \right]\\
l^{k}_{\mathrm{L}}\left(\mathbf{y}\right) = 0, \forall k \in \left[ q_{\mathrm{L}} \right]
\end{matrix} \right\} \\
\Omega_{i}\left(\mathbf{y}, \mathbf{x}_{-i}\right) &=& \left\{\mathbf{x}_{i}\in\mathbb{R}^{n_{i}} \middle|
\begin{matrix}
h^{j}_{i}\left(\mathbf{y}, \mathbf{x}\right) \leq 0, \forall j\in\left[ p_i \right]\\
l^{k}_{i}\left(\mathbf{y}, \mathbf{x}\right) = 0, \forall k\in\left[ q_i \right]
\end{matrix}  \right\}
\end{eqnarray}
Suppose the following four conditions are satisfied: 
(1) $h_{\mathrm{L}}^{j}\left(\mathbf{y}\right)$'s are convex on $\mathbf{y}$; 
(2) $l_{\mathrm{L}}^{k}\left(\mathbf{y}\right)$'s are linear on $\mathbf{y}$; 
(3) $h_{i}^{j}\left(\mathbf{y}, \mathbf{x}\right)$'s are convex on $\mathbf{x}$; and 
(4) $l_{i}^{k}\left(\mathbf{y}, \mathbf{x}\right)$'s are linear on $\mathbf{x}$. Then, we compute the Stackelberg equilibrium of $\Gamma$ in polynomial time with respect to the number of followers. Specifically, the time complexity of computing the Stackelberg equilibrium of $\Gamma$ is $O\left(N^{3.5}\right)$ using Algorithm 1. 

\end{theorem}

\proof{Proof of Theorem \ref{thm5}}
We can convert the $1-N$ generalized Stackelberg game $\Gamma=\left<\mathbf{F}, f_{\mathrm{L}}, \left(f_i\right)_{i\in\mathbf{F}}, \Omega_{\mathrm{L}}, \left(\Omega_i\right)_{i\in\mathbf{F}}\right>$ to the $1-1$ Stackelberg game $\hat{\Gamma}=\left<\left\{1\right\}, f_{\mathrm{L}}, f_{\mathrm{F}}, \Omega_{\mathrm{L}}, \Omega_{\mathrm{F}}\right>$ by Theorem \ref{thm4}. A single update of the leader decision $\mathbf{y}$ in $\hat{\Gamma}$ is conducted by the following three steps:

\begin{itemize}
\item \textbf{Step1. Computing the lower-level solution $\mathbf{w}^{(t)}$ from given $\mathbf{y}^{(t)}$:} Let $n:=\sum_{i\in\mathbf{F}}{n_i}$, $p:=\sum_{i\in\mathbf{F}}{p_i}$, and $q:=\sum_{i\in\mathbf{F}}{q_i}$. Since the gradient of the objective function of follower $i$ is strongly monotone, the iteration complexity of the projected gradient descent is $O(\log{(1/\rho_1)})$ where $\rho_1$ is a residual error of search direction $\mathbf{D} \left(\mathbf{y}^{(t)}, \mathbf{x}^{(k)} \right)$, that is, $\rho_1 \geq ||\mathbf{D} \left( \mathbf{y}^{(t)}, \mathbf{x}^{(k)} \right)||$ \citep{fliege2019complexity}. Next, the gradient update cost per iteration is $O\left(n\right)$. Since $h_{i}^{j}\left(\mathbf{y}, \mathbf{x}\right)$'s are convex and $l_{i}^{k}\left(\mathbf{y}, \mathbf{x}\right)$'s are linear on $\mathbf{x}$, the runtime to find $\rho_2$-approximate projection is $O\left(n(p+2q)^{2.5} \log^2{\left(1/\rho_2\right)} + (p+2q)^{3.5} \log{\left(1/\rho_2\right)} \right)$ \citep{usmanova2021fast}. So, the overall runtime $T_1$ to compute the lower-level solution $\mathbf{w}^{(t)}$ from given $\mathbf{y}^{(t)}$ is $O\left(\log{(1/\rho_1)}\times\left(n (p+2q)^{2.5} \log^2{\left(1/\rho_2\right)} + (p+2q)^{3.5} \log{\left(1/\rho_2\right)}\right)\right)$.

\item \textbf{Step2. Computing the gradient of the leader's objective:} Let $\mathbf{S}$ be the set of the KKT conditions, that is defined as equation (\ref{eqn:S}). Since the number of inequality constraints of $\mathbf{S}$ is $2p$ and the number of equality constraints of $\mathbf{S}$ is $n+p+q$, the number of inequality constraints of $\Omega_{\mathrm{F}}$ is $3p$ and the number of equality constraints of $\Omega_{\mathrm{F}}$ is $n+p+2q$. So, the time complexity to compute the gradient of the follower's decision with respect to the leader's decision
$\frac{d \mathbf{w}^{\left(t\right)}\left(\mathbf{y}^{\left(t\right)}\right)}{d \mathbf{y}} = {M_{\mathrm{FF}}^{\left(t\right)}}^{-1} {M^{\left(t\right)}_{\mathrm{F}}}^{\mathrm{T}} \left({M^{\left(t\right)}_{\mathrm{F}}} {M^{\left(t\right)}_{\mathrm{FF}}}^{-1} {M^{\left(t\right)}_{\mathrm{F}}}^{\mathrm{T}}\right)^{-1}
\left({M^{\left(t\right)}_{\mathrm{F}}} {M^{\left(t\right)}_{\mathrm{FF}}}^{-1} {M^{\left(t\right)}_{\mathrm{LF}}} - {M^{\left(t\right)}_{\mathrm{L}}}\right)-{M^{\left(t\right)}_{\mathrm{FF}}}^{-1} {M^{\left(t\right)}_{\mathrm{LF}}}$ is $O\left(\left(n_{\mathrm{L}}+n+4p+2q\right)^3\right)$ when we use pseudo-inverse. Then, the runtime $T_2$ to compute the gradient of the leader's objective, is expressed in equation (\ref{eqn:grad}), is $O\left(\left(n_{\mathrm{L}}+n+4p+2q\right)^3 + n_{\mathrm{L}}^2 +  n_{\mathrm{L}}^2 n\right)=O\left(\left(n_{\mathrm{L}}+n+4p+2q\right)^3\right)$.

\item \textbf{Step3. Updating the leader's decision $\mathbf{y}^{(t+1)}$:} The time complexity to add the leader's gradient is $O(n_{\mathrm{L}})$. The runtime to find $\rho_3$-approximate projection is $O\left( n_{\mathrm{L}} (p_{\mathrm{L}}+2q_{\mathrm{L}})^{2.5} \log^2{\left(1/\rho_3\right)} + (p_{\mathrm{L}}+2q_{\mathrm{L}})^{3.5} \log{\left(1/\rho_3\right)} \right)$ \cite{usmanova2021fast}. So, the overall runtime $T_3$ to update the leader's decision $\mathbf{y}^{(t+1)}$ is $O\left(n_{\mathrm{L}} (p_{\mathrm{L}}+2q_{\mathrm{L}})^{2.5} \log^2{\left(1/\rho_3\right)} + (p_{\mathrm{L}}+2q_{\mathrm{L}})^{3.5} \log{\left(1/\rho_3\right)} \right)$.

\end{itemize}

As a result, the time complexity of computing the Stackelberg equilibrium of $\Gamma$ is $T=t_{\mathrm{max}}\left(T_1+T_2+T_3\right)$ where the maximum number of iteration of the PIGD algorithm is $t_{\mathrm{max}}$. Since $t_{\mathrm{max}}$, $\rho_1$, $\rho_2$, and $\rho_3$ are constant, and $p$, $q$, and $n$ are linear to $\left|\mathbf{F}\right|=N$, the total runtime $T$ to compute the Stackelberg equilibrium of $\Gamma$ is $O\left(N^{3.5} + \left(N+n_{\mathrm{L}}\right)^3 + n_{\mathrm{L}} \left(p_{\mathrm{L}}+2q_{\mathrm{L}}\right)^{2.5} + \left(p_{\mathrm{L}}+2q_{\mathrm{L}}\right)^{3.5} \right) = O \left( N^{3.5}\right)$, that is polynomial to the number of followers. \Halmos

\endproof

In the following sections, we formulate real-world problems in the form of the $1-N$ Stackelberg games $\Gamma$ and analyze the results of applying our algorithms.

\section{Sharing Platform Problem}
\label{sec:problems}
We employ the generalized Stackelberg game concept and its solution finding algorithm to model the operation of a sharing platform and compute its generalized Stackelberg equilibrium. Particularly, we consider the two problems of deriving operating strategies for EV charging stations:

\begin{itemize}
    \item The first problem is optimizing the one-time charging price for EV users \citep{tushar2012economics} in which a platform operator determines the price of electricity, and EV users determine the optimal amount of charging for their satisfaction. This problem is considered to verify that the proposed algorithm can produce a solution converging to the true equilibrium solution computed analytically. 
    
    \item The second problem is the EV dispatching problem in optimizing the spatially varying charging price for EV users to optimally balance the demand and supply over every charging station \citep{zhou2015optimal} in which a platform operator determines the price of electricity and EV users determine which charging station to go to. The second problem is considered to validate that the proposed modeling and computing method can reliably increase the leader's objective in a more complex problem where there is no known analytical solution. We compare the performance of the proposed PIGD algorithm with the proximal algorithm designed to find a stationary point without considering the hierarchical structure.
\end{itemize}

\subsection{One-time EV Charging Problem}
\label{subsec:problem1}
\subsubsection{Problem description.}
We consider the EV charging problem with one operator and $N$ EVs formulated in \citep{tushar2012economics}. After the operator decides the electricity price $p\in{\mathbb{R}}$ to maximize his profit, each EV $i$ requests a charging amount $x_{i}\in\mathbb{R}$ to maximize his level of satisfaction. In this problem, the followers should satisfy the joint constraint so that the total energy requirement does not exceed the available energy capacity. The $1-N$ generalized Stackelberg game of the one-time EV charging problem is formulated as follows:
\setlength{\arraycolsep}{0.0em}
\begin{eqnarray}
\label{eqn:problem1}
\textrm{Leader's problem :}\quad\max_{p}&&\ p\sum_{i=1}^{N}{x_i}\nonumber\\
\textrm{Follower $i$'s problem :}\quad\max_{x_{i}}&&\ b_{i}x_{i}-\frac{1}{2}s_{i}x_{i}^{2}-px_{i}\nonumber\\
{\rm{s.t.}}&&\ \sum_{j=1}^{N}x_{j}\le C
\end{eqnarray}
where the parameters $b_{i}$, $s_{i}$, and $C$ represent battery capacity, satisfaction parameter, and joint charging limit of EVs, respectively.

\subsubsection{Transformation to $1-1$ Stackelberg game.}
\label{subsubsec:problem1gse}
We transform the one-time EV charging problem into a $1-1$ Stackelberg game having the same solution by applying the proposed converting scheme. By following the transforming procedure described in Appendix \ref{appendix:transform1}, the one-time EV charging problem is given as follows:
\setlength{\arraycolsep}{0.0em}
\begin{eqnarray}
\textrm{Leader's problem :}\quad\max_{p}&&\ p\sum_{i=1}^{N}{x_i}\nonumber\\
\textrm{Followers' joint problem :}\quad\max_{\mathbf{x},\mathbf{z},\mu}&&\ \sum_{i=1}^{N}(-b_{i}+s_{i}x_{i}+p)(z_{i}-x_{i})\nonumber\\
{\rm{s.t.}}&&\ \sum_{i=1}^{N}x_{i}\le C,\ \sum_{i=1}^{N}z_{i}\le C\nonumber\\
&&\ -b_{i}+s_{i}x_{i}+p+\mu=0,\ \forall i\in\left[N\right]\nonumber\\
&&\ \mu\left(\sum_{i}^{N}z_{i}-C\right)=0,\ \mu \geq 0
\label{eqn:D}
\end{eqnarray}
where the last two constraints are obtained from the KKT condition of $\mathbf{z}^{*}=\argmin\limits_{\mathbf{z}\in\prod_{i\in\mathbf{F}}{\Omega_i\left(p, \mathbf{z}_{-i}\right)}}{\mathbf{D}\left(p, \mathbf{x}\right)^{\mathrm{T}}\left(\mathbf{x}-\mathbf{z}\right)}$.

\subsection{EV Dispatching Problem}
\label{subsec:problem2}
\subsubsection{Problem description.}
We reformulate the problem proposed by \citep{zhou2015optimal}, which is the Stackelberg game, into the generalized Stackelberg game by adding the joint constraints that the EVs should satisfy. 

In this EV dispatching problem,  the operators determine the price of electricity for $M$ charging stations distributed in a city, and $N$ EVs distributed over the city choose the charging station to use. The operator wants to regulate the expected number of EVs charging at the station $m$ to be close to the given fixed value $V^m$. For that, the operator sets the electricity price vector $\mathbf{p}=\left\{p^{m}\right\}_{m\in \left[M\right]}$ for the charging stations which is bounded by minimum price $p_{\text{min}}$ and maximum price $p_{\text{max}}$. Station $m$ has an energy limit $L^{m}$ that can be supplied and the maximum number of EVs $U^{m}$ that can be accommodated.

After the operator sets the electricity price, EV $i$ decides which charging station to use for charging a fixed amount of electricity $E_{i}$. The decision of the EV $i$, denoted by $\mathbf{x}_{i}=\left\{x_{i}^{m}\right\}_{m\in\left[M\right]}$, is represented as a probability distribution that EV $i$ chooses the station $m$. Then the expected number of EVs that will charge at the station $m$ is given by $v^{m}=\sum\limits_{i=1}^{N}x_{i}^{m}$. EV $i$ determines its destination to minimize the cost induced by the distance $\mathbf{d}_{i}=\left\{d_{i}^{m}\right\}_{m\in \left[M\right]}$, electricity price $\mathbf{p}$, and congestion level $\mathbf{v}=\left\{v^{m}\right\}_{m\in\left[M\right]}$ of the charging stations. In addition, since the importance of each cost term varies from person to person, each EV has its own coefficients representing the priority of distance, price, and congestion level, denoted by $\alpha_{i}^{d}, \alpha_{i}^ {p}$, and $\alpha_{i}^{v}$ for each EV $i$, respectively.

The $1-N$ generalized Stackelberg game of the EV dispatching problem is formulated as follows:
\setlength{\arraycolsep}{0.0em}
\begin{eqnarray}
\label{eqn:problem2}
\textrm{Leader's problem :}\quad\min_{\mathbf{p}}&&\ \sum_{m=1}^{M}{(v^{m}-V^{m})^{2}}\nonumber\\
{\rm{s.t.}}&&\ p_{\text{min}}\le p^{m} \le p_{\text{max}},\forall{m}\in\left[M\right]\nonumber\\
\textrm{Follower $i$'s problem :}\quad\min_{\mathbf{x}_{i}}&&\ \sum\limits_{m=1}^{M}{\left(\alpha_{i}^{d}d_{i}^{m}x_{i}^{m}+\alpha_{i}^{p}E_{i}p^{m}x_{i}^{m}+\alpha_{i}^{v}x_{i}^{m}v^{m}\right)}\nonumber\\
{\rm{s.t.}}&&\ 0\le x_{i}^{m} \le 1,\ \forall{m}\in\left[M\right]\nonumber\\
&&\ \sum\limits_{m=1}^{M}{x_{i}^{m}}=1\nonumber \\
&&\ \sum_{j=1}^{N}{x_{j}^{m}E_{j}}\le L^{m},\ v^{m}\le U^{m},\ \forall{m}\in\left[M\right]
\end{eqnarray}
The last two inequalities represent the constraints due to the maximum amount of electricity that can be offered and the maximum number of EVs that can be accommodated at the station $m$.

\subsubsection{Transformation to $1-1$ Stackelberg game.}
By following the transforming procedure described in Appendix \ref{appendix:transform2}, the EV dispatching problem is given as follows:
\setlength{\arraycolsep}{0.0em}
\begin{eqnarray}
\textrm{Leader's problem :}\quad\min_{\mathbf{p}}&&\ \sum_{m=1}^{M}{(v^{m}-V^{m})^{2}}\nonumber\\
{\rm{s.t.}}&&\ p_{\text{min}}\le p^{m} \le p_{\text{max}},\forall{m}\in\left[M\right]\nonumber\\
\textrm{Followers' joint problem :}\quad\max_{\mathbf{x},\mathbf{z},\boldsymbol{\mu}, \boldsymbol{\lambda}}&&\ \sum_{i=1}^{N}\sum_{m=1}^{M}\left(A_{i}^{m}+\alpha_{i}^{v}\left(x_{i}^{m}+v^{m}\right)\right)(z_{i}^{m}-x_{i}^{m})\nonumber\\
{\rm{s.t.}}&&\ 0\le x_{i}^{m} \le 1,\ \forall{i}\in\left[N\right],\ \forall{m}\in\left[M\right]\nonumber\\
&&\sum\limits_{m=1}^{M}{x_{i}^{m}}=1, \forall{i}\in\left[N\right]\nonumber\\
&&\sum_{i=1}^{N}{x_{i}^{m}E_{i}}\le L^{m},\ v^{m}\le U^{m},\ \forall{m}\in\left[M\right]\nonumber\\
&&\ 0\le z_{i}^{m} \le 1,\ \forall{i}\in\left[N\right],\ \forall{m}\in\left[M\right]\nonumber\\
&&\sum\limits_{m=1}^{M}{z_{i}^{m}}=1, \forall{i}\in\left[N\right]\nonumber\\
&&\sum_{i=1}^{N}{z_{i}^{m}E_{i}}\le L^{m},\ \sum_{i=1}^{N}{z_{i}^{m}}\le U^{m},\ \forall{m}\in\left[M\right]\nonumber\\
&&A_{i}^{m}+\alpha_{i}^{v}\left(x_{i}^{m}+v^{m}\right)-\mu_{i,m}^{1}+\mu_{i,m}^{2}\nonumber\\
&&\quad\quad+E_{i}\mu_{m}^{3}+\mu_{m}^{4}+\lambda_{i}=0,\ \forall i\in \left[N\right],\ \forall m\in \left[M\right]\nonumber\\
&&\mu_{i,m}^{1}z_{i}^{m}=0,\ \mu_{i,m}^{2}(z_{i}^{m}-1)=0,\ \forall i\in \left[N\right],\ \forall m\in \left[M\right]\nonumber\\
&&\mu_{m}^{3}\left(\sum\limits_{i=1}^{N}\left(z_{i}^{m}E_{i}\right)-L^{m}\right)=0,\ \forall m\in \left[M\right]\nonumber\\ 
&&\mu_{m}^{4}\left(\sum\limits_{i=1}^{N}z_{i}^{m}-U^{m}\right)=0,\ \forall m\in \left[M\right]\nonumber\\
&&\mu_{i,m}^{1},\mu_{i,m}^{2},\mu_{m}^{3},\mu_{m}^{4} \ge 0, \ \forall{i}\in\left[N\right],\ \forall{m}\in\left[M\right]
\end{eqnarray}
where $A_{i}^{m}=\alpha_{i}^{d}d_{i}^{m}+\alpha_{i}^{p}E_{i}p^{m}$. Here, $\boldsymbol{\mu}=\left\{\mu_{i,m}^{1},\mu_{i,m}^{2},\mu_{m}^{3},\mu_{m}^{4}\right\}_{i\in \left[N\right], m\in \left[M\right]}$, and $\boldsymbol{\lambda}=\left\{\lambda_{i}\right\}_{i\in \left[N\right]}$ are Lagrange variables of the inequality constraints and equality constraints, respectively.

Like most generalized Stackelberg games, the EV dispatching problem has no analytic solution or known algorithm for computing it, which hinders the validation of the proposed computing method. Therefore, we consider a baseline pricing strategy induced by the proximal algorithm to compare with the result of the PIGD algorithm. The proximal algorithm iteratively updates the leader's and $N$ followers' decisions until reaching the stationary point. The leader and $N$ followers update their decision by best responding to others' decisions by optimizing a regularized objective function where the penalty term regularizes the changes of their decision compared to the previous step. The proximal algorithm does not consider the hierarchical structure of the problem but guarantees convergence to the generalized Nash equilibrium. The detailed algorithm is provided in Appendix \ref{subsec:algbaseline}.

The proximal algorithm is also used in the one-time EV charging problem as a baseline to show that the result from the proximal algorithm usually does not converge to the generalized Stackelberg equilibrium. In the next section, we discuss the performance of the proposed algorithm with the numerical results.

\section{Simulation Results and Analysis}
We apply the PIGD algorithm to derive the generalized Stackelberg equilibrium of the proposed generalized Stackelberg games and compare it with the analytically found equilibrium and the baseline result. First, we assessed the convergence of the PIGD algorithm with the one-time EV charging problem by comparing the derived equilibrium strategy with the true generalized Stackelberg equilibrium computed in Appendix \ref{subsec:problem1sol}. After that, with the EV dispatching problem, we compared the objective values of both the leader and followers derived by the PIGD algorithm and the proximal algorithm to show that our algorithm performs better than the proximal algorithm from the leader's perspective.

\subsection{Convergence of the Algorithm Using One-time EV Charging Problem}

\begin{figure}[!ht]
    \centering
    \subfigure[Leader's Action]{
    \includegraphics[width=.48\linewidth]{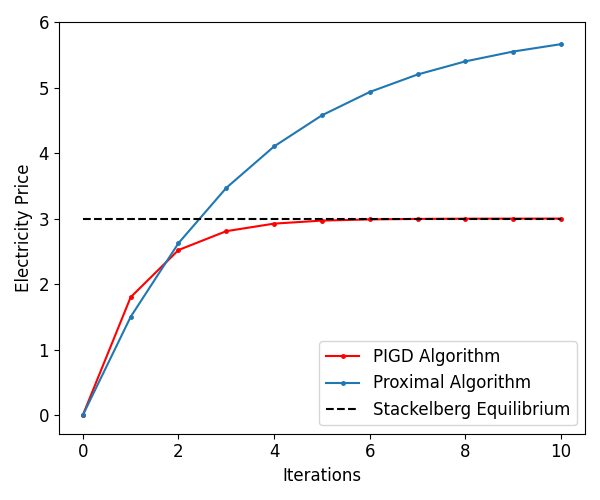}} 
    \subfigure[EV's Action]{
    \includegraphics[width=.48\linewidth]{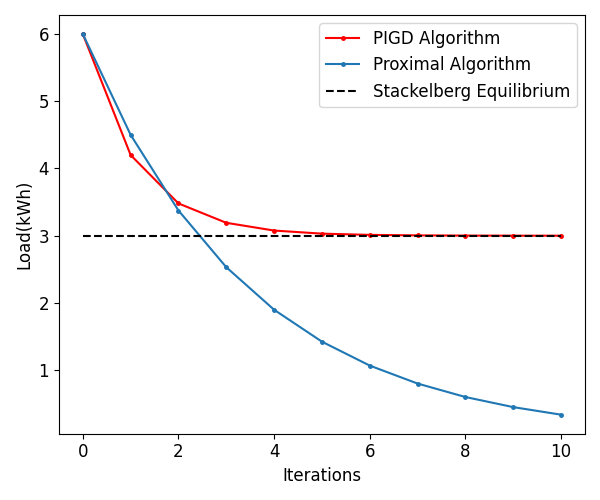}}
    \caption{In (a) and (b), the x-axis represents the number of updates, and the y-axis represents the leader's decision and a single follower's action, respectively.}
    \label{fig:convergence}
\end{figure}

\begin{figure}[!ht]
    \centering
    \subfigure[Leader's Objective]{
    \includegraphics[width=.48\linewidth]{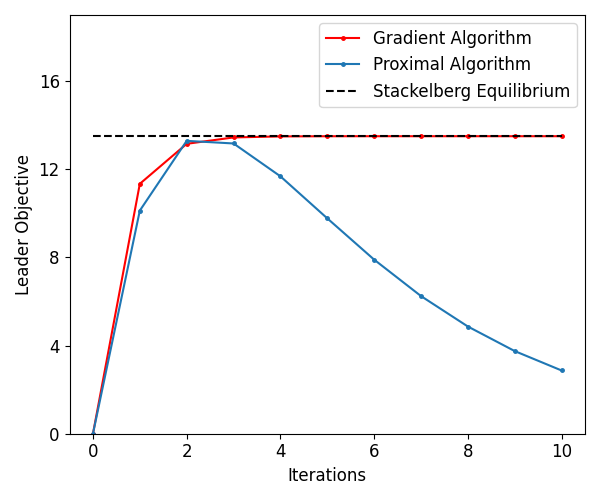}}
    \subfigure[EV's Objective]{
    \includegraphics[width=.48\linewidth]{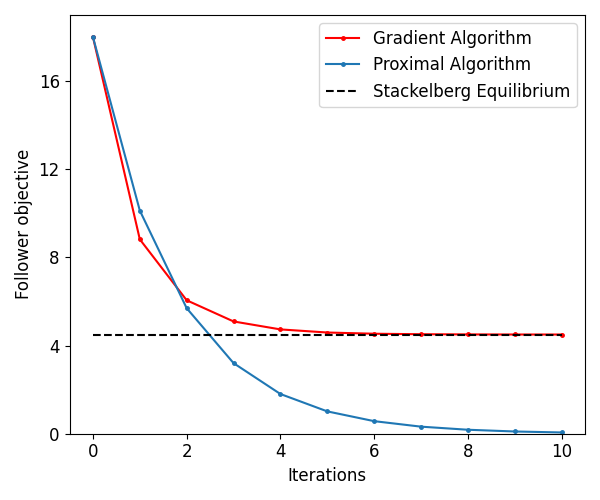}}
    \caption{In (a) and (b), the x-axis represents the number of updates, and the y-axis represents the leader's objective and a single follower's objective, respectively.}
    \label{fig:problem1util}
\end{figure}

We first verify that the PIGD algorithm guarantees the convergence of the solution to the generalized Stackelberg equilibrium by solving the one-time EV charging problem. Figure \ref{fig:convergence} (a) shows how the PIGD algorithm (red line) and the proximal algorithm (blue line) update the leader's decision for each iteration. Figure \ref{fig:convergence} (b) shows the corresponding followers' decision change when the PIGD algorithm and the proximal algorithm are applied, respectively. As shown in the figures, the decision of the leader and the followers converge to the generalized Stackelberg equilibrium computed analytically. The process of computing the analytical solution is provided in Appendix \ref{subsec:problem1sol}. These results imply that the PIGD algorithm can find the generalized Stackelberg equilibrium, while the proximal algorithm fails to produce the generalized Stackelberg equilibrium.

Figure \ref{fig:problem1util} shows how the leader's and followers' objective value changes as the iteration of the PIGD algorithm and the proximal algorithm proceeds. The gradient-based algorithm continuously increases the leader's profit until reaching the generalized Stackelberg equilibrium (dashed line). However, the proximal algorithm fails to converge {to} the generalized Stackelberg equilibrium; it initially reaches the equilibrium but diverges from it. This result shows that the proximal algorithm is not suitable for solving the Stackelberg game because it ignores the hierarchy in which the leader acts first.

\subsection{Efficiency of the Algorithm Using EV Dispatching Problem}
\label{subsec:efficiency}
\begin{figure}[!ht]
    \subfigure[Proximal Algorithm]{
    \includegraphics[width=.49\linewidth]{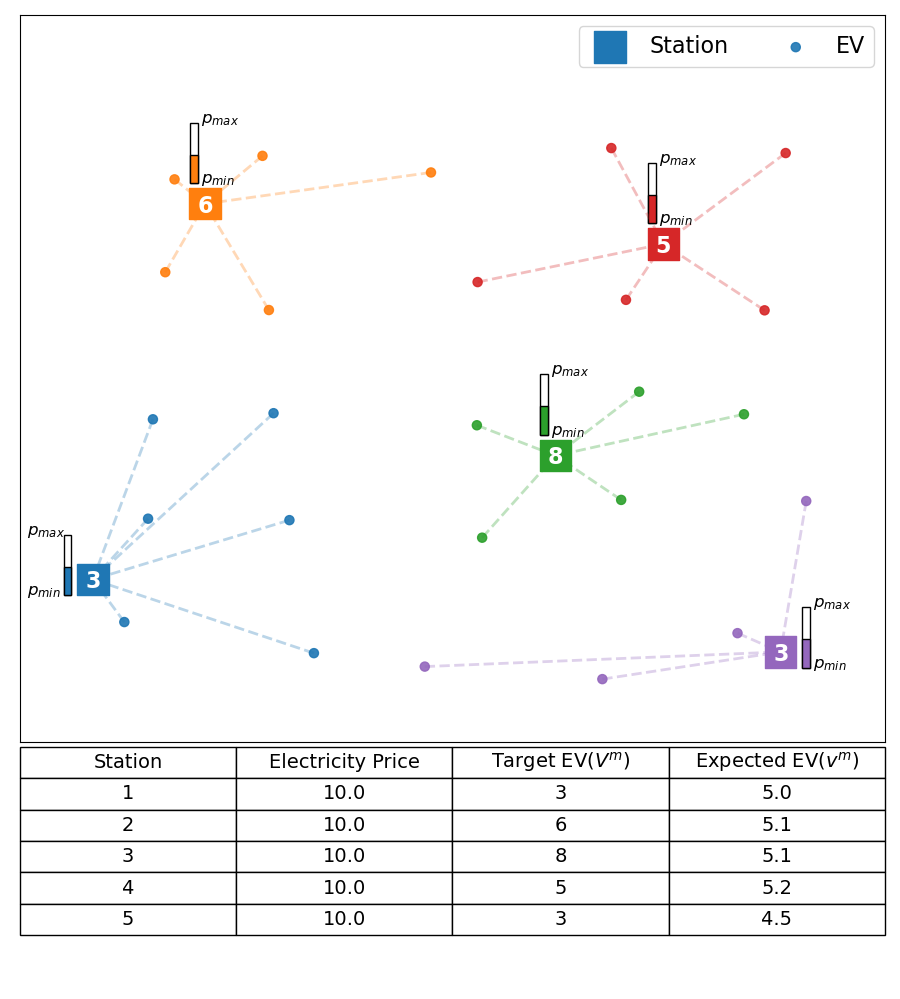}}
    \subfigure[PIGD algorithm]{
    \includegraphics[width=.49\linewidth]{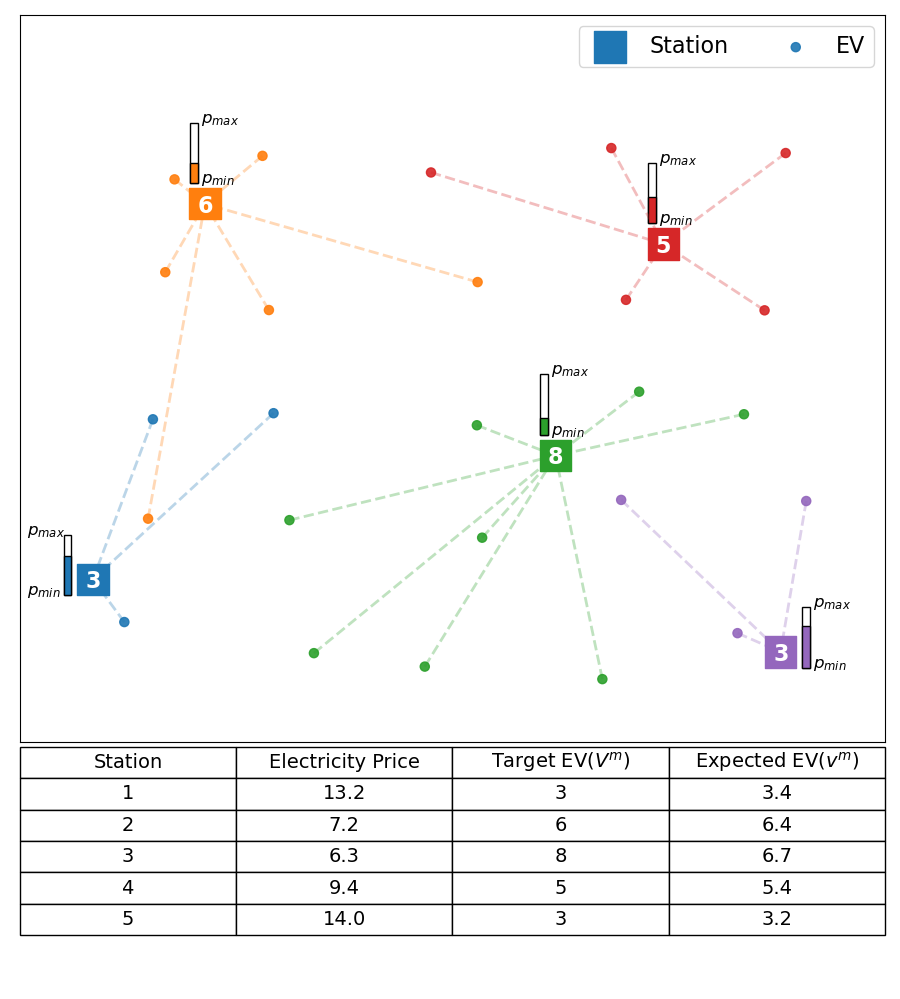}}
    \caption{Dispatched results according to the electricity price induced by the PIGD and the proximal algorithm. The station are numbered from the left. The table shows the electricity price, target EV numbers, and expected EV numbers at each charging station.}
    \label{fig:problem2data4}
\end{figure}

To validate the effectiveness of the proposed algorithm, we solve the EV dispatching problems. In this subsection, we show how PIGD algorithm and the proximal algorithm solve a specific EV charging problem with parameters $N=25$, $M=5$, $p_{\mathrm{max}}=20$, $p_{\mathrm{min}}=1$. The EVs and charging stations are sampled uniformly from the rectangular 2-dimensional space. The require loads $E_{i}$ of the EVs are sampled from uniform distribution on $\left[0.2, 1\right]$ and the priority coefficient $\alpha_{i}^{d},\ \alpha_{i}^{p},\ \alpha_{i}^{v}$ are sampled from the uniform distribution on $\left[0.2, 0.4\right]$. We set the initial electricity price identically for all charging stations. Each algorithm iteratively updates the electricity price to indirectly control the EVs' destination so that the desired capacity is satisfied. 

Figure \ref{fig:problem2data4} shows the problem setting and experimental results of a single problem. Each station shows the target EV number, and the partially colored bar represents the relative electricity price. The PIGD sets the different charging prices for stations to influence an EVs' charging station allocation. It assigns high electricity prices for station $1$ and $5$, whose target numbers are small, and low electricity prices for station $3$, whose target number is high. As a result, more EVs tend to go to station $3$ and fewer EVs to station $1$ and $5$. Please note that the EV allocation is determined not solely by the charging price but also by the distance to the charging stations and the expected waiting time.

On the other hand, the proximal algorithm only induces the same charging price because it cannot model the hierarchical influence of the operator's decision on the EV users. As a result, the EVs determine the charging stations only considering the distance to the charging stations and the expected waiting time. In conclusion, the proposed algorithm can derive an effective operating strategy of the charging station platform while properly considering the hierarchical interactions among the operator and the EV users.

\begin{figure}[!ht]
    \subfigure[Leader]{
    \includegraphics[width=.320\linewidth]{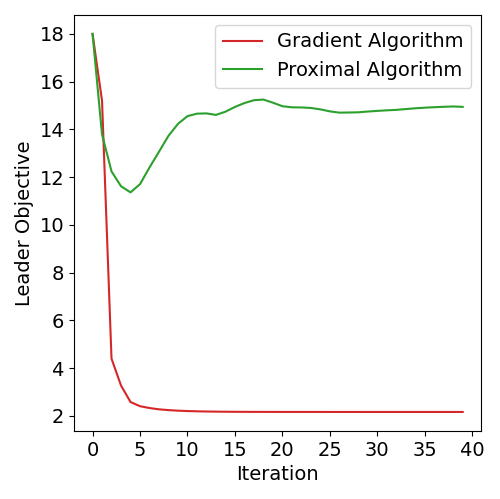}} 
    \subfigure[Individual EVs]{
    \includegraphics[width=.320\linewidth]{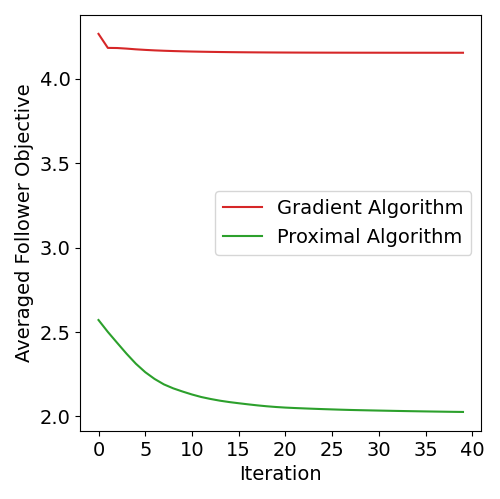}}
    \subfigure[Averaged EVs]{
    \includegraphics[width=.320\linewidth]{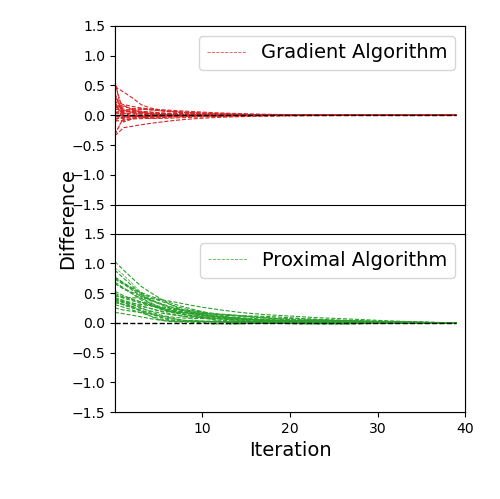}}
    
    \caption{The changes of the leader and the EVs' objective during the updating procedure. The x-axis represents the iterations of each algorithm. In (b), y-axis represent the difference between objective during the iterations and the last iteration.}
    \label{fig:objective}
\end{figure}

In Figure \ref{fig:objective} (b), each dashed line represents the objective value curve of a single EV. We align the ending point of the curves to be zero to show the convergence trend clearly. This shows that all the follower's decisions converge in both algorithms. Figure \ref{fig:objective} (c) shows how the average objective value of EVs varies with the iteration of the two algorithms. The figure shows that both algorithms continuously improve the average objective value (lower is better). It is worth noting that the proximal algorithm induces a lower average objective value of EVs, but it does not indicate the inefficiency of the proposed PIGD algorithm. In this Stackelberg game, the leader takes action first; thus, the leader has the advantage of achieving a better objective value than the followers, which can result in worse objective values for the followers. Therefore, there is no justification that the objective value of the followers of the PIGD algorithm should be better than the value of the proximal algorithm.

We investigate the general performance of the proposed PIGD algorithm by solving the EV dispatching problems with a different number $N$ of EVs and $M$ of charging stations. Mainly, we investigate the varying trend of the objective value of the platform operator for the proposed method with the increases in the problem size (i.e., $N$ and $M$). We sampled a total of 30 problem instances per every combination of $N$ and $M$. 

\begin{table}[!ht]
    \centering
    \begin{tabular}{||c|c||c|c||}
    \hline
        \multicolumn{2}{||c||}{Parameter} & \multicolumn{2}{c||}{Leader Objective} \\ \hline
        M & N & PIGD Algorithm & Proximal Algorithm  \\ \hline\hline
        \multirow{2}{*}{5} & 25 & $\mathbf{3.16}$ & 39.7  \\ \cline{2-4}
         & 50 & $\mathbf{58.2}$ & 190 \\ \hline
        \multirow{2}{*}{10} & 50 & $\mathbf{7.09}$ & 85.6 \\ \cline{2-4}
         & 100 & $\mathbf{78.8}$ & 350  \\ \cline{2-4}
         & 200 & $\mathbf{421}$ & 1103  \\ \hline
        \multirow{3}{*}{20} & 50 & $\mathbf{4.53}$ & 40.1  \\ \cline{2-4}
         & 100 & $\mathbf{20.3}$ & 203  \\ \cline{2-4}
         & 200 & $\mathbf{215}$ & 804  \\ \hline
    \end{tabular}
    \caption{The average of the resulted leader's objective for each parameter.}
    \label{table:performance}
\end{table}

Table \ref{table:performance} compares the leader's objective values achieved by PIGD algorithm and the proximal algorithm. Since the optimal objective value varies depending on sampled problem instances (i.e., the locations of the EVs and stations), only the average objective of the leader is used as a performance measure. As the number $N$ of EVs increase, the objective value of the leader (i.e., the deviation of the target EV assignments and the actual assignment) increases because the platform must manage more EVs. On the contrary, as the number $M$ of charging stations increases, the objective value of the leader tends to decrease because the operator can deploy more various pricing strategies. The results show that PIGD algorithm induces a significantly lower leader's objective value regardless of the problem sizes, indicating that PIGD algorithm can effectively derive the control strategy of an EV charging station operator.

\section{Conclusion}
In this study, we propose the general method to find a generalized Stackelberg equilibrium of the $1-N$ generalized Stackelberg game (single-leader multi-follower game). First, we defined the $1-N$ generalized Stackelberg game and provided the conditions where a generalized Stackelberg equilibrium always exists. Then, we convert the $1-N$ generalized Stackelberg game into the $1-1$ Stackelberg game and apply the PIGD algorithm to compute a generalized Stackelberg equilibrium in polynomial time. The numerical results demonstrate the convergence and effectiveness of our algorithm. The proposed methodology has no restrictions on the problem structure; thus, the proposed modeling and computing methods can be applied to derive an efficient operating strategy for various sharing platforms, such as ride-sharing, car-sharing, and space-sharing. 

\newpage
\bibliography{references}

\begin{thebibliography}{30}
\providecommand{\natexlab}[1]{#1}
\providecommand{\url}[1]{\texttt{#1}}
\providecommand{\urlprefix}{URL }

\bibitem[{Ahmadi \protect\BIBand{} Zhang(2021)}]{ahmadi2021semidefinite}
Ahmadi AA, Zhang J (2021) Semidefinite programming and nash equilibria in
  bimatrix games. \emph{INFORMS Journal on Computing} 33(2):607--628.

\bibitem[{Ba \protect\BIBand{} Pang(2022)}]{ba2022exact}
Ba Q, Pang JS (2022) Exact penalization of generalized nash equilibrium
  problems. \emph{Operations Research} 70(3):1448--1464.

\bibitem[{Bichler et~al.(2023)Bichler, Kohring, \protect\BIBand{}
  Heidekr{\"u}ger}]{bichler2023learning}
Bichler M, Kohring N, Heidekr{\"u}ger S (2023) Learning equilibria in
  asymmetric auction games. \emph{INFORMS Journal on Computing} .

\bibitem[{Chuong(2020)}]{chuong2020optimality}
Chuong TD (2020) Optimality conditions for nonsmooth multiobjective bilevel
  optimization problems. \emph{Annals of Operations Research} 287:617--642.

\bibitem[{Contardo \protect\BIBand{} Sefair(2022)}]{contardo2022progressive}
Contardo C, Sefair JA (2022) A progressive approximation approach for the exact
  solution of sparse large-scale binary interdiction games. \emph{INFORMS
  Journal on Computing} 34(2):890--908.

\bibitem[{Dedzo et~al.(2012)Dedzo, Fotso, \protect\BIBand{}
  Pieume}]{dedzo2012solution}
Dedzo FF, Fotso LP, Pieume CO (2012) Solution concepts and new optimality
  conditions in bilevel multiobjective programming .

\bibitem[{DeMiguel \protect\BIBand{} Xu(2009)}]{demiguel2009stochastic}
DeMiguel V, Xu H (2009) A stochastic multiple-leader stackelberg model:
  analysis, computation, and application. \emph{Operations Research}
  57(5):1220--1235.

\bibitem[{Dempe \protect\BIBand{} Zemkoho(2020)}]{dempe2020bilevel}
Dempe S, Zemkoho A (2020) Bilevel optimization. \emph{Springer optimization and
  its applications} 161.

\bibitem[{Facchinei \protect\BIBand{} Kanzow(2007)}]{facchinei2007generalized}
Facchinei F, Kanzow C (2007) Generalized nash equilibrium problems. \emph{4or}
  5(3):173--210.

\bibitem[{Facchinei \protect\BIBand{} Pang(2003)}]{facchinei2003finite}
Facchinei F, Pang JS (2003) \emph{Finite-dimensional variational inequalities
  and complementarity problems} (Springer).

\bibitem[{Fliege et~al.(2019)Fliege, Vaz, \protect\BIBand{}
  Vicente}]{fliege2019complexity}
Fliege J, Vaz AIF, Vicente LN (2019) Complexity of gradient descent for
  multiobjective optimization. \emph{Optimization Methods and Software}
  34(5):949--959.

\bibitem[{Gould et~al.(2021)Gould, Hartley, \protect\BIBand{}
  Campbell}]{gould2021deep}
Gould S, Hartley R, Campbell D (2021) Deep declarative networks. \emph{IEEE
  Transactions on Pattern Analysis and Machine Intelligence} 44(8):3988--4004.

\bibitem[{Kleinert et~al.(2021)Kleinert, Labb{\'e}, Ljubi{\'c},
  \protect\BIBand{} Schmidt}]{kleinert2021survey}
Kleinert T, Labb{\'e} M, Ljubi{\'c} I, Schmidt M (2021) A survey on
  mixed-integer programming techniques in bilevel optimization. \emph{EURO
  Journal on Computational Optimization} 9:100007.

\bibitem[{Li et~al.(2018)Li, Zhao, Liu, Zhao, Wang, Gooi, Li, \protect\BIBand{}
  Ding}]{li2018interactive}
Li Y, Zhao T, Liu C, Zhao Y, Wang P, Gooi HB, Li K, Ding Z (2018) An
  interactive decision-making model based on energy and reserve for electric
  vehicles and power grid using generalized stackelberg game. \emph{IEEE
  Transactions on Industry Applications} 55(4):3301--3309.

\bibitem[{L{\"u} \protect\BIBand{} Wan(2014)}]{lu2014smoothing}
L{\"u} YB, Wan ZP (2014) A smoothing method for solving bilevel multiobjective
  programming problems. \emph{Journal of the Operations Research Society of
  China} 2(4):511--525.

\bibitem[{Maharjan et~al.(2013)Maharjan, Zhu, Zhang, Gjessing,
  \protect\BIBand{} Basar}]{maharjan2013dependable}
Maharjan S, Zhu Q, Zhang Y, Gjessing S, Basar T (2013) Dependable demand
  response management in the smart grid: A stackelberg game approach.
  \emph{IEEE Transactions on Smart Grid} 4(1):120--132.

\bibitem[{Malitsky(2015)}]{malitsky2015projected}
Malitsky Y (2015) Projected reflected gradient methods for monotone variational
  inequalities. \emph{SIAM Journal on Optimization} 25(1):502--520.

\bibitem[{Mehlitz \protect\BIBand{} Zemkoho(2021)}]{mehlitz2021sufficient}
Mehlitz P, Zemkoho AB (2021) Sufficient optimality conditions in bilevel
  programming. \emph{Mathematics of operations research} 46(4):1573--1598.

\bibitem[{Nash~Jr(1950)}]{Nash1950Equilibrium}
Nash~Jr JF (1950) Equilibrium points in n-person games. \emph{Proceedings of
  the national academy of sciences} 36(1):48--49.

\bibitem[{Qu \protect\BIBand{} Zhao(2013)}]{qu2013methods}
Qu B, Zhao J (2013) Methods for solving generalized nash equilibrium.
  \emph{Journal of Applied Mathematics} 2013:1--6.

\bibitem[{Stackelberg et~al.(1952)}]{Stackelberg1952Theory}
Stackelberg Hv, et~al. (1952) Theory of the market economy .

\bibitem[{Tsaknakis \protect\BIBand{} Hong(2021)}]{tsaknakis2021finding}
Tsaknakis I, Hong M (2021) Finding first-order nash equilibria of zero-sum
  games with the regularized nikaido-isoda function. \emph{International
  Conference on Artificial Intelligence and Statistics}, 1189--1197 (PMLR).

\bibitem[{Tushar et~al.(2012)Tushar, Saad, Poor, \protect\BIBand{}
  Smith}]{tushar2012economics}
Tushar W, Saad W, Poor HV, Smith DB (2012) Economics of electric vehicle
  charging: A game theoretic approach. \emph{IEEE Transactions on Smart Grid}
  3(4):1767--1778.

\bibitem[{Usmanova et~al.(2021)Usmanova, Kamgarpour, Krause, \protect\BIBand{}
  Levy}]{usmanova2021fast}
Usmanova I, Kamgarpour M, Krause A, Levy K (2021) Fast projection onto convex
  smooth constraints. \emph{International Conference on Machine Learning},
  10476--10486 (PMLR).

\bibitem[{Von~Heusinger \protect\BIBand{}
  Kanzow(2009{\natexlab{a}})}]{von2009optimization}
Von~Heusinger A, Kanzow C (2009{\natexlab{a}}) Optimization reformulations of
  the generalized nash equilibrium problem using nikaido-isoda-type functions.
  \emph{Computational Optimization and Applications} 43:353--377.

\bibitem[{Von~Heusinger \protect\BIBand{}
  Kanzow(2009{\natexlab{b}})}]{von2009relaxation}
Von~Heusinger A, Kanzow C (2009{\natexlab{b}}) Relaxation methods for
  generalized nash equilibrium problems with inexact line search. \emph{Journal
  of Optimization Theory and Applications} 143:159--183.

\bibitem[{Wang et~al.(2018)Wang, Hoang, Niyato, Wang, \protect\BIBand{}
  Kim}]{wang2018stackelberg}
Wang W, Hoang DT, Niyato D, Wang P, Kim DI (2018) Stackelberg game for
  distributed time scheduling in rf-powered backscatter cognitive radio
  networks. \emph{IEEE Transactions on Wireless Communications}
  17(8):5606--5622.

\bibitem[{Yoon et~al.(2015)Yoon, Choi, Park, \protect\BIBand{}
  Bahk}]{yoon2015stackelberg}
Yoon SG, Choi YJ, Park JK, Bahk S (2015) Stackelberg-game-based demand response
  for at-home electric vehicle charging. \emph{IEEE Transactions on Vehicular
  Technology} 65(6):4172--4184.

\bibitem[{Zhang et~al.(2010)Zhang, Qu, \protect\BIBand{} Xiu}]{zhang2010some}
Zhang J, Qu B, Xiu N (2010) Some projection-like methods for the generalized
  nash equilibria. \emph{Computational optimization and applications}
  45(1):89--109.

\bibitem[{Zhou et~al.(2015)Zhou, Liu, Yang, \protect\BIBand{}
  Guan}]{zhou2015optimal}
Zhou H, Liu C, Yang B, Guan X (2015) Optimal dispatch of electric taxis and
  price making of charging stations using stackelberg game. \emph{IECON
  2015-41st Annual Conference of the IEEE Industrial Electronics Society},
  004929--004934 (IEEE).

\end{thebibliography}

\newpage
\begin{APPENDICES}
\section{Proof}\label{A}
\subsection{Proof of Theorem \ref{thm3}}\label{A.1}
Let $\mathrm{G}\left(\mathbf{y}\right) = \left<\mathbf{F}, \left( f_{i} \right)_{i \in \mathbf{F}}, \left(\Omega_{i}\right)_{i\in\mathbf{F}}\right>$ be the $N$ followers' generalized normal-form game when the leader's decision is $\mathbf{y}$. Since the $1-N$ generalized Stackelberg game $\Gamma=\left<\mathbf{F}, f_{\mathrm{L}}, \left(f_{i}\right)_{i\in\mathbf{F}}, \Omega_{\mathrm{L}}, \left(\Omega_{i}\right)_{i\in\mathbf{F}}\right>$ satisfies the condition 1, 2, and 4 of Theorem \ref{thm3}, there is the unique variational equilibrium of $\mathrm{G}\left(\mathbf{y}\right)$ for all $\mathbf{y}\in\Omega_{\mathrm{L}}$ by Theorem \ref{thm1}. Then, $\Gamma$ has a variational Stackelberg equilibrium $\left(\mathbf{y}^{*}, \mathbf{x}^{*}\left(\mathbf{y}^{*}\right)\right)$ by Definition \ref{def3}.

Because the $1-N$ generalized Stackelberg game $\Gamma$ satisfies the condition 1, 2, and 3 of Theorem \ref{thm3}, the unique variational equilibrium $\mathbf{x}^{*}\left(\mathbf{y}\right)$ of $\mathrm{G}\left(\mathbf{y}\right)$ is also a generalized Nash equilibrium of $\mathrm{G}\left(\mathbf{y}\right)$ for all $\mathbf{y}\in\Omega_{\mathrm{L}}$ by Theorem \ref{thm2}. Then, the following equation holds:
\setlength{\arraycolsep}{0.0em}
\begin{eqnarray}
\sup_{\mathbf{x}\left(\mathbf{y}^{*}\right)\in \mathrm{GNE}\left(\mathbf{y}^{*}\right)}f_{\mathrm{L}}\left(\mathbf{y}^{*}, \mathbf{x}\left(\mathbf{y}^{*}\right)\right)&\geq&\sup_{\mathbf{x}\left(\mathbf{y}^{*}\right)\in \mathrm{VE}\left(\mathbf{y}^{*}\right)}f_{\mathrm{L}}\left(\mathbf{y}^{*}, \mathbf{x}\left(\mathbf{y}^{*}\right)\right)\nonumber\\
&=&f_{\mathrm{L}}\left(\mathbf{y}^{*},\mathbf{x}^{*}\left(\mathbf{y}^{*}\right)\right)
\label{eqn:appendix1}
\end{eqnarray}
By equation (\ref{eqn:def3}) of Definition \ref{def3}, 
\setlength{\arraycolsep}{0.0em}
\begin{eqnarray}
f_{\mathrm{L}}\left(\mathbf{y}^{*},\mathbf{x}^{*}\left(\mathbf{y}^{*}\right)\right)&\geq&f_{\mathrm{L}}\left(\mathbf{y},\mathbf{x}^{*}\left(\mathbf{y}\right)\right)\nonumber\\
&\geq&\inf_{\mathbf{x}\left(\mathbf{y}\right)\in \mathrm{GNE}\left(\mathbf{y}\right)}f_{\mathrm{L}}\left(\mathbf{y}, \mathbf{x}\left(\mathbf{y}\right)\right), \forall \mathbf{y} \in \Omega_{\mathrm{L}}
\label{eqn:appendix2}
\end{eqnarray}
Thus, equation (\ref{eqn:def4}) of Definition \ref{def4} holds for the variational Stackelberg equilibrium $\left(\mathbf{y}^{*}, \mathbf{x}^{*}\left(\mathbf{y}^{*}\right)\right)$ by equations (\ref{eqn:appendix1}) and (\ref{eqn:appendix2}). Therefore, the variational Stackelberg equilibrium of $\Gamma=\left<\mathbf{F}, f_{\mathrm{L}}, \left(f_{i}\right)_{i\in\mathbf{F}}, \Omega_{\mathrm{L}}, \left(\Omega_{i}\right)_{i\in\mathbf{F}}\right>$ is also a generalized Stackelberg equilibrium $\Gamma$. \Halmos

\subsection{Proof of Proposition \ref{prop1}}\label{A.2}

Let $\mathbf{x}^{*}\left(\mathbf{y}\right)$ is a generalized Nash equilibrium of the followers' generalized normal-form game $\mathrm{G}\left(\mathbf{y}\right) = \left<\mathbf{F}, \left( f_{i} \right)_{i \in \mathbf{F}}, \left(\Omega_{i}\right)_{i\in\mathbf{F}}\right>$, and $\left(\mathbf{y}^{*}, \mathbf{x}^{*}\left(\mathbf{y}^{*}\right)\right)$ is a generalized Stackelberg equilibrium of the $1-N$ generalized Stackelberg game $\Gamma=\left<\mathbf{F}, f_{\mathrm{L}}, \left(f_{i}\right)_{i\in\mathbf{F}}, \Omega_{\mathrm{L}}, \left(\Omega_{i}\right)_{i\in\mathbf{F}}\right>$. Since the set of a generalized Nash equilibrium of $\mathrm{G}\left(\mathbf{y}\right)$ is unique for all $\mathbf{y}\in\Omega_{\mathrm{L}}$, a generalized Stackelberg equilibrium $\left(\mathbf{y}^{*}, \mathbf{x}^{*}\left(\mathbf{y}^{*}\right)\right)$ satisfies the following equation:
\setlength{\arraycolsep}{0.0em}
\begin{eqnarray}
f_{\mathrm{L}}\left(\mathbf{y}^{*}, \mathbf{x}^{*}\left(\mathbf{y}^{*}\right)\right) \geq f_{\mathrm{L}}\left(\mathbf{y}, \mathbf{x}^{*}\left(\mathbf{y}\right)\right), \forall \mathbf{y} \in \Omega_{\mathrm{L}}
\label{eqn:appendixA2}
\end{eqnarray}

Assume that $\Gamma$ has more than one generalized Stackelberg equilibrium. That is, there are two generalized Stackelberg equilibrium $\left(\mathbf{y}^{\mathrm{1}}, \mathbf{x}^{*}\left(\mathbf{y}^{\mathrm{1}}\right)\right)$ and $\left(\mathbf{y}^{\mathrm{2}}, \mathbf{x}^{*}\left(\mathbf{y}^{\mathrm{2}}\right)\right)$ which satisfy equation (\ref{eqn:appendixA2}). Since $\Omega_{\mathrm{L}}$ is closed and convex, $\mathbf{y}^{\mathrm{3}}:=\alpha\mathbf{y}^{\mathrm{1}}+\left(1-\alpha\right)\mathbf{y}^{\mathrm{2}}\in\Omega_{\mathrm{L}}$ for all $\alpha \in \left[0, 1\right]$. Moreover, $f_{\mathrm{L}}\left(\mathbf{y}^{\mathrm{3}}, \mathbf{x}^{*}\left(\mathbf{y}^{\mathrm{3}}\right)\right) \geq \alpha f_{\mathrm{L}}\left(\mathbf{y}^{\mathrm{1}}, \mathbf{x}^{*}\left(\mathbf{y}^{\mathrm{1}}\right)\right) + \left(1-\alpha\right) f_{\mathrm{L}}\left(\mathbf{y}^{\mathrm{2}}, \mathbf{x}^{*}\left(\mathbf{y}^{\mathrm{2}}\right)\right)$ because $f_{\mathrm{L}}\left(\mathbf{y}, \mathbf{x}\left(\mathbf{y}\right)\right)$ is strictly concave on $\mathbf{y}\in\Omega_{\mathrm{L}}$, that is $f_{\mathrm{L}}\left(\mathbf{y}^{\mathrm{3}}, \mathbf{x}^{*}\left(\mathbf{y}^{\mathrm{3}}\right)\right) \geq f_{\mathrm{L}}\left(\mathbf{y}^{\mathrm{1}}, \mathbf{x}^{*}\left(\mathbf{y}^{\mathrm{1}}\right)\right)$ or $f_{\mathrm{L}}\left(\mathbf{y}^{\mathrm{3}}, \mathbf{x}^{*}\left(\mathbf{y}^{\mathrm{3}}\right)\right) \geq f_{\mathrm{L}}\left(\mathbf{y}^{\mathrm{2}}, \mathbf{x}^{*}\left(\mathbf{y}^{\mathrm{2}}\right)\right)$. It contradicts that $\left(\mathbf{y}^{\mathrm{2}}, \mathbf{x}^{*}\left(\mathbf{y}^{\mathrm{2}}\right)\right)$ and $\left(\mathbf{y}^{\mathrm{2}}, \mathbf{x}^{*}\left(\mathbf{y}^{\mathrm{2}}\right)\right)$ satisfy equation (\ref{eqn:appendixA2}). Thus, $1-N$ generalized Stackelberg game $\Gamma=\left<\mathbf{F}, f_{\mathrm{L}}, \left(f_{i}\right)_{i\in\mathbf{F}}, \Omega_{\mathrm{L}}, \left(\Omega_{i}\right)_{i\in\mathbf{F}}\right>$ has the unique generalized Stackelberg equilibrium. \Halmos

\newpage

\section{Formula Derivation}\label{B}
\subsection{Transformation of $1-N$ generalized Stackelberg game to $1-1$ Stackelberg game}
\subsubsection{One-time EV charging problem}
\label{appendix:transform1}
\hfill\\
The one-time EV charging problem is formulated as follows:
\setlength{\arraycolsep}{0.0em}
\begin{eqnarray}
\textrm{Leader's problem :}\quad\max_{p}&&\ p\sum_{i=1}^{N}{x_i}\nonumber\\
\textrm{Follower $i$'s problem :}\quad\max_{x_{i}}&&\ b_{i}x_{i}-\frac{1}{2}s_{i}x_{i}^{2}-px_{i}\nonumber\\
{\rm{s.t.}}&&\ \sum_{j=1}^{N}x_{j}\le C
\end{eqnarray}
With the given leader decision $p$, the variational inequality problem of the followers' subgame is finding $\mathbf{x}\in\prod\limits_{i=1}^{N}\Omega_{i}(p,\mathbf{x}_{-i})=\left\{\mathbf{x}\,\middle|\,\sum\limits_{i=1}^{N}x_{i}\le C\right\}$ which satisfies the following equation:
\setlength{\arraycolsep}{0.0em}
\begin{eqnarray}
\label{eqn:vi_origin}
\sum_{i=1}^{N}(-b_{i}+s_{i}x_{i}+p)(z_{i}-x_{i}) \ge 0,\quad \forall \mathbf{z}\in\prod\limits_{i=1}^{N}\Omega_{i}(p,\mathbf{z}_{-i})=\left\{\mathbf{z}\,\middle|\,\sum\limits_{i=1}^{N}z_{i}\le C\right\}
\end{eqnarray}
where $\mathbf{x}$ and $\mathbf{z}$ represent the vector of the followers' action.

Instead of solving the variational inequality problem directly, we change it to a bilevel problem with $\mathbf{x}$ and $\mathbf{z}$ as variables of each level. Then the $1-N$ generalized Stackelberg game is converted to a three-level optimization problem as follows:
\setlength{\arraycolsep}{0.0em}
\begin{eqnarray}
\label{eqn:vi}
\textrm{Leader's problem :}\quad\max_{p}&&\ p\sum_{i=1}^{N}{x_i}\nonumber\\
\textrm{Followers' upper-level problem :}\quad\max_{\mathbf{x}}&&\ \sum_{i=1}^{N}(-b_{i}+s_{i}x_{i}+p)(z_{i}-x_{i})\nonumber\\
{\rm{s.t.}}&&\ \sum_{i=1}^{N}x_{i}\le C\nonumber\\
\textrm{Followers' lower-level problem :}\quad\min_{\mathbf{z}}&&\ \sum_{i=1}^{N}(-b_{i}+s_{i}x_{i}+p)(z_{i}-x_{i})\nonumber\\
{\rm{s.t.}}&&\ \sum_{i=1}^{N}z_{i}\le C
\end{eqnarray}
In equation (\ref{eqn:vi}), the optimal condition of the lower-level problem can be substituted with the KKT condition and added as constraints at the upper-level problem. By adding a new Lagrange variable $\mu$, the three-level optimization problem is converted to $1-1$ Stackelberg game as follows:
\setlength{\arraycolsep}{0.0em}
\begin{eqnarray}
\textrm{Leader's problem :}\quad\max_{p}&&\ p\sum_{i=1}^{N}{x_i}\nonumber\\
\textrm{Followers' joint problem :}\quad\max_{\mathbf{x},\mathbf{z},\mu}&&\ \sum_{i=1}^{N}(-b_{i}+s_{i}x_{i}+p)(z_{i}-x_{i})\nonumber\\
{\rm{s.t.}}&&\ \sum_{i=1}^{N}x_{i}\le C,\ \sum_{i=1}^{N}z_{i}\le C\nonumber\\
&&\ -b_{i}+s_{i}x_{i}+p+\mu=0,\ i\in\left[N\right]\nonumber\\
&&\ \mu\left(\sum_{i=1}^{N}z_{i}-C\right)=0,\ \mu \geq 0
\end{eqnarray}

\subsubsection{EV dispatching problem}
\label{appendix:transform2}
\hfill\\
The EV dispatching problem is formulated as follows:
\setlength{\arraycolsep}{0.0em}
\begin{eqnarray}
\textrm{Leader's problem :}\quad\min_{\mathbf{p}}&&\ \sum_{m=1}^{M}{(v^{m}-V^{m})^{2}}\nonumber\\
{\rm{s.t.}}&&\ p_{\text{min}}\le p^{m} \le p_{\text{max}},\forall{m}\in\left[M\right]\nonumber\\
\textrm{Follower $i$'s problem :}\quad\min_{\mathbf{x}_{i}}&&\ \sum\limits_{m=1}^{M}{\left(\alpha_{i}^{d}d_{i}^{m}x_{i}^{m}+\alpha_{i}^{p}E_{i}p^{m}x_{i}^{m}+\alpha_{i}^{v}x_{i}^{m}v^{m}\right)}\nonumber\\
{\rm{s.t.}}&&\ 0\le x_{i}^{m} \le 1,\ \forall{m}\in\left[M\right]\nonumber\\
&&\sum\limits_{m=1}^{M}{x_{i}^{m}}=1\nonumber\\
&&\sum_{j=1}^{N}{x_{j}^{m}E_{j}}\le L^{m},\ v^{m}\le U^{m},\ \forall{m}\in\left[M\right]
\end{eqnarray}

When the leader's decision $\mathbf{p}$ is given, the definition of the variational inequality problem of the $N$ followers' subgame is finding the followers' decision $\mathbf{x}\in\prod\limits_{i=1}^{N}\Omega_{i}\left(\mathbf{p},\mathbf{x}_{-i}\right)$ which satisfies the following equation:
\setlength{\arraycolsep}{0.0em}
\begin{eqnarray}
\sum_{i=1}^{N}\sum_{m=1}^{M}\left(A_{i}^{m}+\alpha_{i}^{v}\left(x_{i}^{m}+v^{m}\right)\right)(z_{i}-x_{i}) \ge 0,\quad \forall \mathbf{z}\in\prod\limits_{i=1}^{N}\Omega_{i}(\mathbf{p},\mathbf{z}_{-i})
\end{eqnarray}
where $A_{i}^{m}:=\alpha_{i}^{d}d_{i}^{m}+\alpha_{i}^{p}E_{i}p^{m}$ and $\mathbf{x}$, $\mathbf{z}$ represent the vector of the followers' action.

Instead of solving the variational inequality problem directly, we change it to a bilevel problem with $\mathbf{x}$ and $\mathbf{z}$ as variables of each level. Then the $1-N$ generalized Stackelberg game is converted to a three-level optimization problem as follows:
\setlength{\arraycolsep}{0.0em}
\begin{eqnarray}
\label{eqn:vi_2}
\textrm{Leader's problem :}\quad\max_{\mathbf{p}}&&\ \sum_{m=1}^{M}{(v^{m}-V^{m})^{2}}\nonumber\\
{\rm{s.t.}}&&\ p_{\text{min}}\le p^{m} \le p_{\text{max}},\forall{m}\in\left[M\right]\nonumber\\
\textrm{Followers' upper-level problem :}\quad\max_{\mathbf{x}}&&\ \sum_{i=1}^{N}\sum_{m=1}^{M}\left(A_{i}^{m}+\alpha_{i}^{v}\left(x_{i}^{m}+v^{m}\right)\right)(z_{i}-x_{i})\nonumber\\
{\rm{s.t.}}&&\ \mathbf{x}\in\prod\limits_{i=1}^{N}\Omega_{i}(\mathbf{p},\mathbf{x}_{-i})\nonumber\\
\textrm{Followers' lower-level problem :}\quad\min_{\mathbf{z}}&&\ \sum_{i=1}^{N}\sum_{m=1}^{M}\left(A_{i}^{m}+\alpha_{i}^{v}\left(x_{i}^{m}+v^{m}·\right)\right)(z_{i}-x_{i})\nonumber\\
{\rm{s.t.}}&&\ \mathbf{z}\in\prod\limits_{i=1}^{N}\Omega_{i}(\mathbf{p},\mathbf{z}_{-i})
\end{eqnarray}
In equation (\ref{eqn:vi_2}), the optimal condition of the lower-level problem can be substituted with the KKT condition and added as constraints at the upper-level problem. By adding new Lagrange variables $\boldsymbol{\mu}=\left\{\mu_{i,m}^{1},\mu_{i,m}^{2},\mu_{m}^{3},\mu_{m}^{4}\right\}_{i\in\left[N\right], m\in\left[M\right]}$ for the inequality constraints and $\boldsymbol{\lambda}=\left\{\lambda_{i}\right\}_{i\in\left[N\right]}$ for the equality constraints, the three-level optimization problem is converted to a $1-1$ Stackelberg game as follows:

\setlength{\arraycolsep}{0.0em}
\begin{eqnarray}
\textrm{Leader's problem :}\quad\min_{\mathbf{p}}&&\ \sum_{m=1}^{M}{(v^{m}-V^{m})^{2}}\nonumber\\
{\rm{s.t.}}&&\ p_{\text{min}}\le p^{m} \le p_{\text{max}},\forall{m}\in\left[M\right]\nonumber\\
\textrm{Followers' joint problem :}\quad\max_{\mathbf{x},\mathbf{z},\boldsymbol{\mu}, \boldsymbol{\lambda}}&&\ \sum_{i=1}^{N}\sum_{m=1}^{M}\left(A_{i}^{m}+\alpha_{i}^{v}\left(x_{i}^{m}+v^{m}\right)\right)(z_{i}^{m}-x_{i}^{m})\nonumber\\
{\rm{s.t.}}&&\ 0\le x_{i}^{m} \le 1,\ \forall{i}\in\left[N\right],\ \forall{m}\in\left[M\right]\nonumber\\
&&\sum\limits_{m=1}^{M}{x_{i}^{m}}=1, \forall{i}\in\left[N\right]\nonumber\\
&&\sum_{i=1}^{N}{x_{i}^{m}E_{i}}\le L^{m},\ v^{m}\le U^{m},\ \forall{m}\in\left[M\right]\nonumber\\
&&\ 0\le z_{i}^{m} \le 1,\ \forall{i}\in\left[N\right],\ \forall{m}\in\left[M\right]\nonumber\\
&&\sum\limits_{m=1}^{M}{z_{i}^{m}}=1, \forall{i}\in\left[N\right]\nonumber\\
&&\sum_{i=1}^{N}{z_{i}^{m}E_{i}}\le L^{m},\ \sum_{i=1}^{N}{z_{i}^{m}}\le U^{m},\ \forall{m}\in\left[M\right]\nonumber\\
&&A_{i}^{m}+\alpha_{i}^{v}\left(x_{i}^{m}+v^{m}\right)-\mu_{i,m}^{1}+\mu_{i,m}^{2}\nonumber\\
&&\quad\quad+E_{n}\mu_{m}^{3}+\mu_{m}^{4}+\lambda_{n}=0,\ \forall n\in \left[N\right],\ \forall m\in \left[M\right]\nonumber\\
&&\mu_{i,m}^{1}z_{i}^{m}=0,\ \mu_{i,m}^{2}(z_{i}^{m}-1)=0,\ \forall i\in \left[N\right],\ \forall m\in \left[M\right]\nonumber\\
&&\mu_{m}^{3}\left(\sum\limits_{i=1}^{N}\left(z_{i}^{m}E_{i}\right)-L^{m}\right)=0,\ \forall m\in \left[M\right]\nonumber\\ &&\mu_{m}^{4}\left(\sum\limits_{i=1}^{N}z_{i}^{m}-U^{m}\right)=0,\ \forall m\in \left[M\right]\nonumber\\
&&\mu_{i,m}^{1},\mu_{i,m}^{2},\mu_{m}^{3},\mu_{m}^{4} \ge 0, \ \forall{i}\in\left[N\right],\ \forall{m}\in\left[M\right]
\end{eqnarray}

\subsection{Generalized Stackelberg Equilibrium of the One-time EV Charging Problem}
\label{subsec:problem1sol}
To compute the generalized Stackelberg equilibrium, we first solve the followers' subgame with the fixed leader's decision $p$. We utilize the KKT condition to compute the variational equilibrium of the followers' subgame. The Lagrangian coefficient of the followers' joint constraint is identical for all followers at the variational equilibrium, which is proven in Theorem 9 \citep{facchinei2007generalized}. We denote this Lagrangian coefficient of joint constraint of the followers' subgame in the one-time EV charging problem as $\mu$. Then the KKT condition of the followers' subgame is derived as follows:
\setlength{\arraycolsep}{0.0em}
\begin{eqnarray}
&&b_{i}-s_{i}x_{i}-p-\mu=0,\ \forall{i}\in\left[N\right]\nonumber\\
&&\sum_{i=1}^{N}x_{i}\le C,\ \mu \left(\sum_{i=1}^{N}x_{i}-C\right)=0, \ \mu\ge0
\label{eqn:prob1kkt}
\end{eqnarray}
We can compute the solution of equation (\ref{eqn:prob1kkt}) by dividing the case of $\mu=0$ and the case of $\sum\limits_{i=1}^{N}x_{i}-C=0$. Then the variational equilibrium of the followers with the given leader decision $p$ is given as follows:
\begin{eqnarray}
x_{i}=
\left\{ 
  \begin{array}{ c l }
    \frac{b_{i}-p}{s_{i}} & \quad \textrm{if } B-pS \le C \\
    \frac{1}{s_{i}}\left(b_{i}-\frac{B-C}{S}\right) & \quad \textrm{if } B-pS > C
  \end{array}
\right.
\quad \mathrm{where}\  B=\sum_{i=1}^{N}\left(\frac{b_{i}}{s_{i}}\right)\  \mathrm{and}\  S=\sum_{i=1}^{N}\left(\frac{1}{s_{i}}\right)
\end{eqnarray}
Here $B$ and $S$ are constant since they only depend on the EVs' battery capacity and satisfaction parameter. Let's assume that $B\le 2C$. When $B-pS \le C$, $x_{i}=\frac{b_{i}-p}{s_{i}}$ is a variational equilibrium point of the followers. Substituting this into the leader's objective $p\sum\limits_{i=1}^{N}x_{i}$, we get a concave quadratic expression for $p$ which has an optimal solution $p^{*}=\frac{B}{2S}$. It also satisfies the condition $B-p^{*}S=\frac{B}{2}\le C$ by the assumption. The optimal solution of the follower $i$ is $x_{i}^{*}=\frac{b_{i}-p^{*}}{s_{i}}$.

\section{Baseline Algorithms}\label{C}
\label{subsec:algbaseline}
\begin{algorithm}
\caption{Proximal Algorithm for $1-N$ Generalized Stackelberg Game $\Gamma$}
    \begin{algorithmic}[1]
        \Require{Feasible leader decision $\mathbf{y}^{(0)}$ and followers' decision $\mathbf{x}^{(0)}\in\Omega_{\mathbf{F}}(\mathbf{y})$, updating coefficients $\tau^{t}>0$} tending to $0$ and  stopping criterion $\epsilon>0$.
        \Repeat \ for $t=0,1,2,\dots$
            \State Compute the leader's next decision by optimizing the proximal function
                \begin{eqnarray}
                    \mathbf{y}^{(t+1)}=\argmax_{\mathbf{y}\in\Omega_{\mathrm{L}}( \mathbf{x}^{(t)})}f_{\mathrm{L}}(\mathbf{y},\mathbf{x})-\frac{\tau^{t}}{2}\|\mathbf{y}^{(t+1)}-\mathbf{y}^{(t)}\|^{2}
                    \label{eqn:proximal}
                \end{eqnarray}
            \For{$i=1,2,\dots,N$}
                \State Compute the follower $i$'s next decision by optimizing the proximal function
                \begin{eqnarray}
                    \mathbf{x}_{i}^{(t+1)}=\argmax_{\mathbf{x}_{i}\in\Omega_{i}(\mathbf{y}^{(t+1)}, \mathbf{x}^{(t)})}f_{i}(\mathbf{y}^{(t+1)}, \mathbf{x}_{i}, \mathbf{x}_{-i}^{(t)})-\frac{\tau^{(t)}}{2}\|\mathbf{x}_{i}^{(t+1)}-\mathbf{x}_{i}^{(t)}\|^{2}
                \end{eqnarray}        
            \EndFor
        \Until{$\left\|\left(\mathbf{y}^{(t+1)},\mathbf{x}^{(t+1)}\right)-\left(\mathbf{y}^{(t)},\mathbf{x}^{(t)}\right)\right\|<\epsilon$}
    \end{algorithmic}
\label{alg:3}
\end{algorithm}
\end{APPENDICES}

\end{document}